\documentclass[aps,prd,preprint,nofootinbib,amsmath,amssymb,superscriptaddress]{revtex4-1}
 \usepackage{amsmath}
\usepackage{amssymb}
\usepackage{graphicx}
\usepackage{color}
\usepackage{bm}
\usepackage{times}
\usepackage{hyperref}
\hypersetup{
  colorlinks=true,
  citecolor=blue,
  linkcolor=blue,
  urlcolor=blue}

\usepackage{epsfig}% Include figure files
\usepackage{dcolumn}% Align table columns on decimal point
\usepackage{float}
\usepackage{multirow}

\usepackage{epsfig}% Include figure files
\usepackage{dcolumn}% Align table columns on decimal point
\usepackage[normalem]{ulem}
\def\be{\begin{equation}}
\def\ee{\end{equation}}
\def\bea{\begin{eqnarray}}
\def\eea{\end{eqnarray}}
\def\gsim{\ \rlap{\raise 2pt\hbox{$>$}}{\lower 2pt \hbox{$\sim$}}\ }
\def\lsim{\ \rlap{\raise 2pt\hbox{$<$}}{\lower 2pt \hbox{$\sim$}}\ }
\def\dslash{\kern-4pt \not{\hbox{\kern-2pt $\partial$}}}
\def\pslash{\not{\hbox{\kern-2pt p}}}

\def\pmue{{{P_{\mu e}} }}

\newcommand{\dcp}{\delta_{\rm CP}}

\newcommand{\mudar}{$\mu$-DAR}

\newcommand{\anumu}{\ensuremath{\bar{\nu}_\mu}}

\newcommand{\epsee}{\varepsilon^{\mu e}_{ee}}

\newcommand{\epsme}{\varepsilon^{\mu e}_{\mu e}}
\newcommand{\epsmm}{\varepsilon^{\mu e}_{\mu\mu}}
\newcommand{\epsmt}{\varepsilon^{\mu e}_{\mu\tau}}

\newcommand{\phime}{\phi^{\mu e}_{\mu e}}
\newcommand{\phimm}{\phi^{\mu e}_{\mu\mu}}

\newcommand{\dm}[1]{\Delta m^2_{#1}}
\newcommand{\re}[1]{\textrm{Re}(#1)}

\begin{document}

\DeclareGraphicsExtensions{.eps,.ps}

\title{Probing muonic charged current nonstandard interactions at decay-at-rest facilities in conjunction with T2HK}

\author{Soumya C.}
\email{soumya.c@iopb.res.in}
\affiliation{ Institute of Physics, Sachivalaya Marg, Sainik School Post, Bhubaneswar 751005, India}

\author{Monojit Ghosh}
\email{manojit@kth.se}
\affiliation{Department of Physics, School of Engineering Sciences,
KTH Royal Institute of Technology, AlbaNova University Center, Roslagstullsbacken 21, SE--106 91 Stockholm, Sweden}
\affiliation{The Oskar Klein Centre for Cosmoparticle Physics, AlbaNova University Center, Roslagstullsbacken 21, SE--106 91 Stockholm, Sweden}

\author{Sushant K. Raut}
\email{sushant@ibs.re.kr}
\affiliation{ Center for Theoretical Physics of the Universe, Institute for Basic Science (IBS), Daejeon, 34051, Korea}
\affiliation{Department of Physics, KAIST, Daejeon 34141, Korea}

\author{Nita Sinha}
\email{nita@imsc.res.in}
\affiliation{The Institute of Mathematical Sciences, Chennai 600113, India}

\author{Poonam Mehta}
\email{pm@jnu.ac.in}
\affiliation{ School of Physical Sciences, Jawaharlal Nehru University,  New Delhi 110067, India}

%\maketitle
\begin{abstract}

The muon decay-at-rest ($\mu$-DAR) facility provides us with an ideal platform to probe purely  muonic charged-current nonstandard neutrino
 interactions (NSIs). We propose to probe this class of NSI effects using antineutrinos from a $\mu$-DAR source in conjunction with neutrinos from the future Tokai to Kamioka  superbeam experiment with megaton Hyper Kamiokande detector (T2HK). Even though muonic NSIs are absent in neutrino production at T2HK, we show that our proposed hybrid setup comprising $\mu$-DAR and T2HK helps in alleviating the parameter degeneracies that can arise in data. 
Analytic considerations reveal that the oscillation probability is most sensitive to the NSI parameter in the $\mu$-e sector. 
For this parameter, we show that 
the $\mu$-DAR setup can improve on the existing bounds down to around 0.01, especially when the data are combined with neutrino data from  T2HK experiment due to the lifting of parameter degeneracies. The high precision with which $\mu$-DAR can measure $\delta_{\rm{CP}}$ is shown to be robust even in the presence of the considered  NSIs. Finally, we show that the combination of  $\mu$-DAR along with T2HK can also be used to put mild constraints on the NSI phase in the vicinity of the maximal CP-violating value for the chosen benchmark value of $\epsme=0.01$. 
\end{abstract}

\maketitle

\section{Introduction}
\label{sec:intro}

Neutrino physics has entered a precision era with a smooth transition (over the past few decades) of the goals, from   measuring precisely the parameters of the neutrino mixing matrix to looking for signals of physics beyond the Standard Model (SM).  While most of oscillation parameters in the standard three flavor paradigm have been measured successfully to varying degrees of precision (for latest global fit to neutrino data, see~\cite{Esteban:2016qun}),  one needs to  measure the value of $\dcp$, figure out the correct octant of $\theta_{23}$ and address the issue of neutrino mass hierarchy. 
In our quest for physics beyond the SM, 
 some of the currently ongoing and future neutrino experiments are expected to play pivotal role  in constraining the new physics parameter space. While effects due to new physics are more intensively looked for at the neutrino experiments, collider experiments allow for a  complementary probe of new physics~\cite{Davidson:2011kr}.

Many directions have been explored to study  new physics  in the neutrino sector such as sterile neutrinos, nonstandard interactions (NSIs),  large extra dimensions, nonunitarity, neutrino decays, violation of charge conjugation, parity and time reversal (CPT) and Lorentz symmetry, quantum decoherence, etc. (see~\cite{Dev:2019anc} and references therein). 
Of these, NSIs of neutrinos is one of the most widely studied new physics scenarios, an idea which originated in an important paper by Wolfenstein~\cite{Wolfenstein:1977ue}  (for reviews, see Refs.~\cite{Miranda:2015dra,Ohlsson:2012kf,Farzan:2017xzy,Dev:2019anc}). Some of the other motivations for NSIs include electroweak leptogenesis~\cite{Pilaftsis:2005rv}, neutrino magnetic moment~\cite{BARBIERI1990251,PhysRevLett.63.228}, neutrino condensate as dark energy~\cite{CHOUDHURY1990113} and   direct detection of dark matter~\cite{Gonzalez-Garcia:2018dep}.
Typically, in the general effective field theory  description, the effective Lagrangian can be of the form of charged current (CC) or neutral current (NC) interaction which can alter production, detection and propagation of neutrinos and lead to rich phenomenology.  Detailed analyses pertaining to the constraints on these NSI parameters have been summarized in~\cite{Davidson:2003ha,Biggio:2009nt,Farzan:2017xzy}. Propagation NSI has been the topic of much recent interest at long baseline accelerator experiments such as Deep Underground Neutrino Experiment (DUNE) (e.g., ~\cite{Masud:2015xva, Coloma:2015kiu,deGouvea:2015ndi,Blennow:2016etl,Deepthi:2017gxg,Masud:2018pig}) as well as reactor (e.g., ~\cite{Kopp:2007ne}), solar (e.g., ~\cite{Das:2010sd}) and atmospheric (e.g., ~\cite{Chatterjee:2014gxa,Choubey:2015xha})~experiments.  While the treatment of propagation NSI does not depend upon the details of particular source or detector type, it should be noted that description of  NSI at source or detector is governed by 
the details of the  production or detection mechanism at play.

In terrestrial experiments, neutrino production occurs via either beta decay or pion decay processes. 
Typically,  the interactions of neutrinos at source and detector are  CC interactions involving both leptons and quarks.  The impact of these NSIs on neutrino oscillation phenomenology has been studied extensively, for example, in the context of  ESS$\nu$SB~\cite{Blennow:2015nxa}, DUNE~\cite{Blennow:2016etl,Bakhti:2016gic},  astrophysical neutrinos~\cite{Blennow:2009rp}, neutrino factory~\cite{Meloni:2009cg}, explaining data from MiniBoone and liquid scintillator neutrino detector (LSND)~\cite{Akhmedov:2010vy} as well as  solar and reactor neutrino experiments~\cite{Bolanos:2008km,Leitner:2011aa,Khan:2013hva,Girardi:2014kca,Khan:2014zwa,Agarwalla:2014bsa,Khan:2017oxw}.
In addition, neutrinos may be also produced via muon-decay process and it is worthwhile to investigate impact of NSIs on processes involving muons  especially in light of the various anomalies (involving muons) such as muon $g-2$ or results involving $B$-meson decays~\cite{Bennett:2006fi,Albrecht:2018vsa}. 

Clean measurement of the CP-violating phase $\dcp$ is a formidable task especially in the context of superbeam experiments~\cite{Rout:2017udo} and it was proposed that experiments with a muon decay-at-rest (\mudar) source could help accomplish this~\cite{Alonso:2010fs,Evslin:2015pya,Agarwalla:2010nn}. Another proposed experiment for clean measurement of the CP phase is called the MuOn-decay MEdium baseline NeuTrino beam facility (MOMENT) which has neutrino beam of around 300 MeV and high flux~\cite{Cao:2014bea}. In a typical \mudar\ experiment, $\anumu$ produced from muon decay (with energy around a few $10$s of MeV) oscillates into $\bar \nu_{e}$ and is detected through the inverse beta-decay (IBD) process. It should be noted that in this setup, not only the flux is very well known but also  the IBD detection cross section of these neutrinos is large and well-measured. Also,  
 the much smaller systematic uncertainties coupled with large detection efficiency and fewer backgrounds allow for a clean measurement of $\dcp$.  In a recent study,  it is proposed that the \mudar\ setup is useful in addressing the question of neutrino mass hierarchy~\cite{Agarwalla:2017nld}.

New physics in the context of \mudar\ has been studied in~\cite{Ge:2016xya,Ge:2016dlx,Ge:2018uhz}.  New physics effects have been studied  in the context of the  MOMENT in~\cite{Tang:2017qen}.
% perhaps refer to https://arxiv.org/pdf/1401.8125.pdf
 In  the present work, we  attempt to constrain the ``CC muonic NSI parameters" at the neutrino production stage using the \mudar\ setup. We also demonstrate the impact of combining data from the future Tokai to Kamioka with megaton Hyper Kamiokande detector (T2HK) superbeam experiment~\cite{Abe:2016ero} which aids in lifting parameter degeneracies~\cite{Ghosh:2015ena} even if CC muonic NSI parameters do not play any role in case of neutrino production in T2HK. The CC NSIs at the detection stage involving quarks have been neglected in the present work.  In general, one must also consider NC NSIs that affect the propagation of neutrinos through matter. However, such effects are energy dependent, similar to the Mikheyev-Smirnov-Wolfenstein (MSW) matter effect. 
Therefore, at low energies, one may assume them to be negligible.

The outline of this paper is as follows. In Sec.~\ref{sec:mudar}, we describe in detail the specifications of the experiments, \mudar\  and T2HK   considered in the present work.  Section~\ref{sec:formalism} contains the formalism for CC NSIs  for the case of muon decay with a brief discussion  on  the existing constraints on  relevant parameters. In Sec.~\ref{sec:results}, we present our results and discuss the outcome. Finally, we end with concluding remarks in Sec.~\ref{sec:conc}.

\section{Experimental details : \mudar\ facility and T2HK}
\label{sec:mudar}

\begin{figure*}[htb!]
\begin{center}
\includegraphics[width=0.5\textwidth]{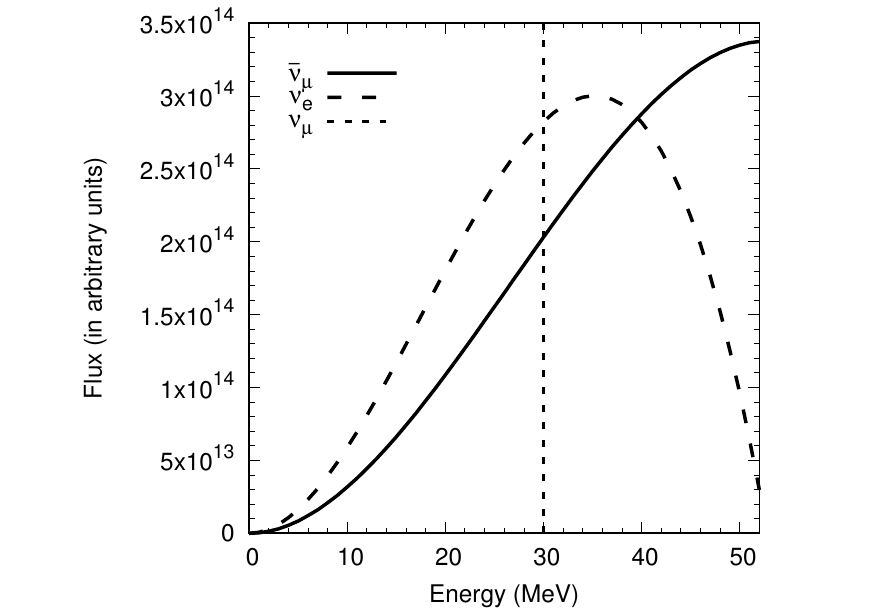}
\end{center}
\caption{\footnotesize{The neutrino spectrum produced by $\pi^+$ and $\mu^+$ decay at rest. The vertical dotted line shows the $\nu_\mu$ from the two body decay  of the pions, $\pi^+ \to \mu^+ + \nu_\mu$. The dashed and solid lines depict the $\nu_e$ and $\bar \nu_\mu $ from the three body decay process.}}
\label{fig0}
\end{figure*}

For the \mudar\ facility, we use the configuration described in Ref.~\cite{Evslin:2015pya}. The \mudar\ source  collides low energy protons into a large enough target, thereby, creating charged pions, $\pi^+$ and $\pi^-$ which stop in the target. Of these, $\pi^-$ may  either get absorbed or decay into $\nu_\mu$ and $\mu^-$ that are absorbed in a reasonably high-$Z$ target. While, the  $\pi^+$ decay at rest into $\mu^+$ and $\nu_\mu$. The $\mu^+$ thus produced stops and decays at rest, producing $e^+$, $\bar{\nu}_\mu$ and $\nu_e$. 
The spectra of neutrinos obtained via the process
\begin{eqnarray}
\pi^+ &\to& \mu^+ + \nu_\mu
\nonumber\\
 && 
 {\LARGE{\hookrightarrow}}   ~~e^+ + \bar \nu_\mu +\nu_e
 % {\HUGE{\rotatebox[origin=c]{180}{$\Lsh$} }}
 \end{eqnarray}
  is depicted in Fig.~\ref{fig0}. 
 The \mudar\ spectrum is known precisely.  The \mudar\ experiment  is well suited to study $\bar{\nu}_\mu \rightarrow \bar{\nu}_e$ oscillations. As the $\bar \nu_e$  have  energies around $30 - 50$ MeV, these can be detected via the IBD process with precisely known cross section. The proposal is to place the \mudar\ accelerator complex in the southern hills of Toyama about  15 km from the site of the Super-Kamiokande (SK) detector and about  23 km from the Hyper-Kamiokande (HK) detector. The SK detector is a 22.5 kt water Cerenkov detector 
and the proposed HK detector consists of two water Cerenkov detectors, each having mass of 187 kt.

The experimental project comprising neutrinos produced at the J-PARC accelerator and detected by the SK and HK detectors is known as the T2HK project~\cite{Abe:2016ero}. In T2HK, both the SK and HK detectors will be located at a distance of 295 km from the J-PARC accelerator site.  
    
As mentioned  in the introduction, our main goal is to constrain the relevant source NSI parameters at \mudar\ facility.  It should be noted that simultaneous operation of \mudar\ and T2HK is very much feasible since the low energy $\bar \nu_e$ can be distinguished at SK and HK from pulsed higher energy neutrinos from J-PARC. The beam-on backgrounds (which are predominantly the intrinsic $\bar \nu_e$ events in the beam) are small at \mudar\ and hence have been neglected in the present study. 
However, we have incorporated the uncertainties in estimating the   background from low energy atmospheric neutrinos~\cite{Evslin:2015pya}.
%
%However we do consider the low energy atmospheric backgrounds based on the discussion in Ref.~\cite{Evslin:2015pya}.} 
% 
We also explore the impact of adding the data from T2HK on our results. In Table \ref{tab:specs}, we have given the details of the  \mudar\ setup~\cite{Evslin:2015pya} and T2HK~\cite{Abe:2016ero}.

%-----------------------------------------------------------
\begin{table*}[htb]
 \begin{tabular}{|p{0.14\textwidth}|p{0.1\textwidth}|p{0.1\textwidth}|p{0.2\textwidth}|p{0.32\textwidth}|}
  \hline
  Experiment & Detector mass (kt) & Baseline (km) & Total p.o.t. & Systematic errors \\
    \hline
  \mudar\ & 22.5(SK) + 374(HK) & 15(SK) + 23(HK) & $1.1 \times 10^{25}$ ($\bar{\nu}$)  & 5\% for both sg and bg (same for both SK and HK). \\
  \hline
   T2HK & 374 & 295 & $27 \times 10^{21}$ ($\nu$) OR $6.75 \times 10^{21}$ ($\nu$) + $20.25 \times 10^{21}$ ($\overline{\nu}$) & $\nu$: 3.3\% for app and disap, $\bar{\nu}$: 6.2\% (4.5\%) for app (disap). Errors are same for sg and bg. \\
  \hline
 \end{tabular}
\caption{Summary of experimental details assumed in our simulations. Here app=appearance channel, disap=disappearance channel, $\nu$=neutrino, $\bar{\nu}$=antineutrino, sg=signal, and bg=background.}
\label{tab:specs}
\end{table*}
%-------------------------------------------------------------------

\section{NSI in the context of \mudar}
\label{sec:formalism}

In the SM, the CC interactions of neutrinos are flavor diagonal by definition. The inclusion of NSI effects can alter this, so that the neutrino produced in association with the charged lepton $\ell_\alpha$ need not be purely $\nu_\alpha$ (as in the case of SI) but can also have an admixture of other flavor $\nu_\beta$ proportional to the (subdominant) NSI term $\varepsilon^{\mu e}_{\alpha\beta}$. Thus, up to a normalization factor, we have  
\[
 \nu^s_\alpha = \sum_\beta (\delta_{\alpha\beta} + \varepsilon^{\mu e}_{\alpha\beta}) \nu_\beta ~. 
\]
The superscript $s$ stands for production NSIs at the source, while the label $\mu e$ on the NSI parameters indicates that these neutrinos are produced in muon decay in conjunction with a muon and an electron. The effective operator responsible for muon decay is $\sim G_F \ \bar{e} \gamma_\rho P_L \nu_e^s \ \bar{\nu}_\mu^s \gamma^\rho P_L \mu$, which modifies the muon-decay neutrino flux as~\cite{Antusch:2006vwa} 
\[
 \Phi^{NSI} = \Phi^{SM} \left( 1+2\re{\epsee} \right) \left( 1+2\re{\epsmm} \right) ~.
\]
Two neutrinos of unknown flavor are produced, of which the antineutrino $\bar{\nu}_\mu^s$ is detected after oscillation while the neutrino $\nu_e^s$ escapes undetected. 
Therefore the probability of $\bar{\nu}^s_\mu \to \bar{\nu}_\rho$ is
\begin{eqnarray}
  P(\bar{\nu}^s_\mu \to \bar{\nu}_\rho) & = & \sum_X P(\bar{\nu}^s_\mu \nu^s_e \to \bar{\nu}_\rho \nu_X)  \nonumber \\
  & = & \sum_X |\mathcal{A}(\bar{\nu}^s_\mu \to \bar{\nu}_\rho ) \mathcal{A}( \nu^s_e \to \nu_X)|^2 \textrm{ \ \ \ \ (assuming uncorrelated propagation)}\nonumber \\
  & = & \sum_X \left| \sum_{\beta,\alpha} (\delta_{\mu\beta} + \varepsilon^{\mu e *}_{\mu\beta}) \bar{\mathcal{A}}_{\beta \rho} (\delta_{e\alpha} + \varepsilon^{\mu e}_{e\alpha}) \mathcal{A}_{\alpha X} \right|^2 \nonumber \\
  & = & \sum_X \left[ \sum_{\beta,\alpha} (\delta_{\mu\beta} + \varepsilon^{\mu e *}_{\mu\beta}) \bar{\mathcal{A}}_{\beta \rho} (\delta_{e\alpha} + \varepsilon^{\mu e}_{e\alpha}) \mathcal{A}_{\alpha X} \right] \left[ \sum_{\delta,\gamma} (\delta_{\mu\delta} + \varepsilon^{\mu e}_{\mu\delta}) \bar{\mathcal{A}}^*_{\delta \rho} (\delta_{e\gamma} + \varepsilon^{\mu e *}_{e\gamma}) \mathcal{A}^*_{\gamma X} \right] ~, \nonumber
 \label{eq:pmeformal}
\end{eqnarray}
where $\mathcal{A}_{\alpha\beta}$ (without/with bar) denotes the standard amplitude for the process $\nu_\alpha \to \nu_\beta$ (for neutrinos/antineutrinos). From the completeness of neutrino states, we have $ \sum_X \mathcal{A}_{\alpha X} \mathcal{A}^*_{\gamma X} = \delta_{\alpha\gamma}$. Using this, and $\sum_\alpha (\delta_{e\alpha} + \varepsilon^{\mu e}_{e\alpha}) (\delta_{e\alpha} + \varepsilon^{\mu e *}_{e\alpha}) \approx ( 1+2\re{\epsee})$ we get
\begin{eqnarray}
 P(\bar{\nu}^s_\mu \to \bar{\nu}_\rho) & = & \left( 1+2\re{\epsee} \right) \left| \sum_\beta (\delta_{\mu\beta} + \varepsilon^{\mu e *}_{\mu\beta}) \bar{\mathcal{A}}_{\beta \rho} \right|^2 \nonumber \\
 & \approx & (1+2\re{\epsee}) \left[ (1+2\re{\epsmm}) \bar{P}_{\mu \rho} + 2\re{\epsme \bar{\mathcal{A}}_{e\rho} \bar{\mathcal{A}}_{\mu\rho}} + 2\re{\epsmt \bar{\mathcal{A}}_{\tau\rho} \bar{\mathcal{A}}_{\mu\rho} } \right] ~,
\end{eqnarray}
where we have only retained terms up to first order in the $\varepsilon^{\mu e}_{\alpha\beta}$ parameters.
It is easy to check that 
$P(\bar{\nu}^s_\mu \to \bar{\nu}_e) + P(\bar{\nu}^s_\mu \to \bar{\nu}_\mu) + P(\bar{\nu}^s_\mu \to \bar{\nu}_\tau) = (1+2\re{\epsee})(1+2\re{\epsmm})$; therefore, the neutrino states in the presence of NSIs (and hence the probability) should be appropriately normalized. These are the same factors $(1+2\re{\epsee})(1+2\re{\epsmm})$ that appear in the modified neutrino flux. Instead of normalizing the probability {and} multiplying the same factor to the SM flux, we choose to do neither with the understanding that these factors cancel out in the calculation of event rates~\cite{Antusch:2006vwa}. Notice that only the parameters $\epsme$, $\epsmm$ and $\epsmt$ are relevant in the normalized probability formula.

The formalism described above is completely general and  is applicable to any muon-decay neutrino source where one of the neutrinos goes undetected. In the specific case of \mudar\ facility, the only relevant oscillation channel is $P(\bar{\nu}^s_\mu \to \bar{\nu}_e)$, since the low energy neutrino cannot produce a $\mu$ or $\tau$ lepton in the detector. Therefore, the relevant formula for the probability (before normalization) is 
\begin{equation}
 P(\bar{\nu}^s_\mu \to \bar{\nu}_e) = (1+2\re{\epsee}) \left[ (1+2\re{\epsmm}) \bar{P}_{\mu e} + 2\re{\epsme \bar{\mathcal{A}}_{ee} \bar{\mathcal{A}}_{\mu e}} + 2\re{\epsmt \bar{\mathcal{A}}_{\tau e} \bar{\mathcal{A}}_{\mu e} } \right] ~.
\end{equation}

An analytical formula for the oscillation probability in the presence of NSIs can be calculated perturbatively in the small parameters, $\alpha=\dm{21}/\dm{31} \approx 0.03$, $\sin\theta_{13} \approx 0.15$ and $\varepsilon^{\mu e}_{\alpha\beta}$~\cite{Kopp:2007ne}. Each NSI term $\varepsilon^{\mu e}_{\alpha\beta}$ can be complex and therefore introduces two new real parameters -- a magnitude $|\varepsilon^{\mu e}_{\alpha\beta}|$ and the corresponding phase $\phi^{\mu e}_{\alpha\beta}$. In general, the formula also depends on the MSW matter potential, $A$. However, since we will restrict our discussion to experiments with short baselines and low energies, it suffices to consider the vacuum oscillation formula and study the role of NSI parameters. Up to first order in the NSI parameters, we have 
\begin{eqnarray}
 P^{NSI} \equiv P(\nu^s_\mu \to \nu_e) & = & 4 \sin^2\theta_{13} \sin^2\theta_{23} \sin^2\Delta \nonumber \\
 & + & 2\alpha\sin\theta_{13} \sin 2\theta_{12} \sin 2\theta_{23} \cos(\Delta+\dcp) \Delta \sin\Delta \nonumber \\
 & - & 4 |\epsme| \sin\theta_{13} \sin\theta_{23} \sin(\Delta+\dcp+\phime) \sin\Delta \nonumber \\
 & - & 2 |\epsme| \alpha \sin 2\theta_{12} \cos\theta_{23} \sin\phime \Delta ~,
 \label{eq:pme}
\end{eqnarray}
where $\Delta = \dm{31}L/4E$. For antineutrinos (which are relevant for our study), we make the replacements $\dcp\to-\dcp$ and $\phime\to-\phime$. Since $\sin\theta_{13}>\alpha,|\epsme|$, the first term which is proportional to $\sin^2\theta_{13}$ dominates the probability. The second and third terms are typically smaller in magnitude, while the fourth term proportional to $\alpha|\epsme|$ is the smallest.

Only the last two terms of the formula show the effect of NSIs. Up to lowest order, the formula involves $\epsme$ but the parameters $\epsmm$ and $\epsmt$ do not appear. Therefore, we expect the sensitivity of the $\pmue$ channel to be better for $\epsme$ than the other two parameters. Apart from modulating the amplitude of the probability spectrum, the NSI also introduces an additional source of CP violation through the phase.  We will refer back to this formula to explain features of the experimental sensitivity.  However, it should be noted that all the  numerical results presented in the present work are exact. 

The $\varepsilon^{\mu e}_{\alpha \beta}$ can be constrained from (i) tests of lepton universality, i.e., by comparing the Fermi constant from muonic and electronic processes, and (ii) the nonobservation of zero-distance flavor conversion at experiments like KARMEN~\cite{Eitel:2000by} and NOMAD~\cite{Astier:2003gs}. This analysis was carried out in Ref.~\cite{Biggio:2009nt} (assuming one nonzero NSI parameter at a time) and the following $90\%$ C.L. bounds were obtained on the magnitudes of the NSIs: 
\begin{equation} 
|\varepsilon^{\mu e}_{\alpha \beta}| < \left(
\begin{array}{ccc}
 0.025 & 0.030 & 0.030 \\
 0.025 & 0.030 & 0.030 \\
 0.025 & 0.030 & 0.030
\end{array} 
\right) 
\end{equation}
In our study, we will constrain the parameters ($\epsme$, $\epsmm$, and $\epsmt$) using the \mudar\ setup and compare them to their existing bounds (0.025, 0.030, 0.030, respectively).

 \begin{table}[h]
\centering
\begin{tabular}{ |l  | c | c |}
\hline
&&\\
Oscillation parameter & True value & Marginalization range \\
&&\\
\hline
%&&\\
%{\sl{SI}} &&\\
&&\\
$\theta_{12}$ [$^\circ$] & $33.5$  &  - \\
$\theta_{13}$ [$^\circ$] & $8.5$ & -\\
$\theta_{23}$ [$^\circ$] & $45$  & $40 - 50$ \\
$\dm{21}$ [$\textrm{eV}^2$]  & $7.5 \times 10^{-5}$ & - \\
$\dm{31}$ (NH) [$\textrm{eV}^2$] & $+2.45 \times 10^{-3}$ & - \\
$\dm{31}$ (IH) [$\textrm{eV}^2$] & $-2.46 \times 10^{-3}$ &  \\
$\dcp$ [$^\circ$] & $-90$ & $-180 - 180$\\
\hline
\end{tabular}
\caption{ Oscillation parameters in the standard three flavor paradigm used in our study~\cite{Esteban:2016qun}. }
\label{tab:parameters}
\end{table}
%-------------------------------------------

\section{Results}
\label{sec:results}

\begin{figure*}[htb!]
\begin{center}
\includegraphics[width=0.98\textwidth]{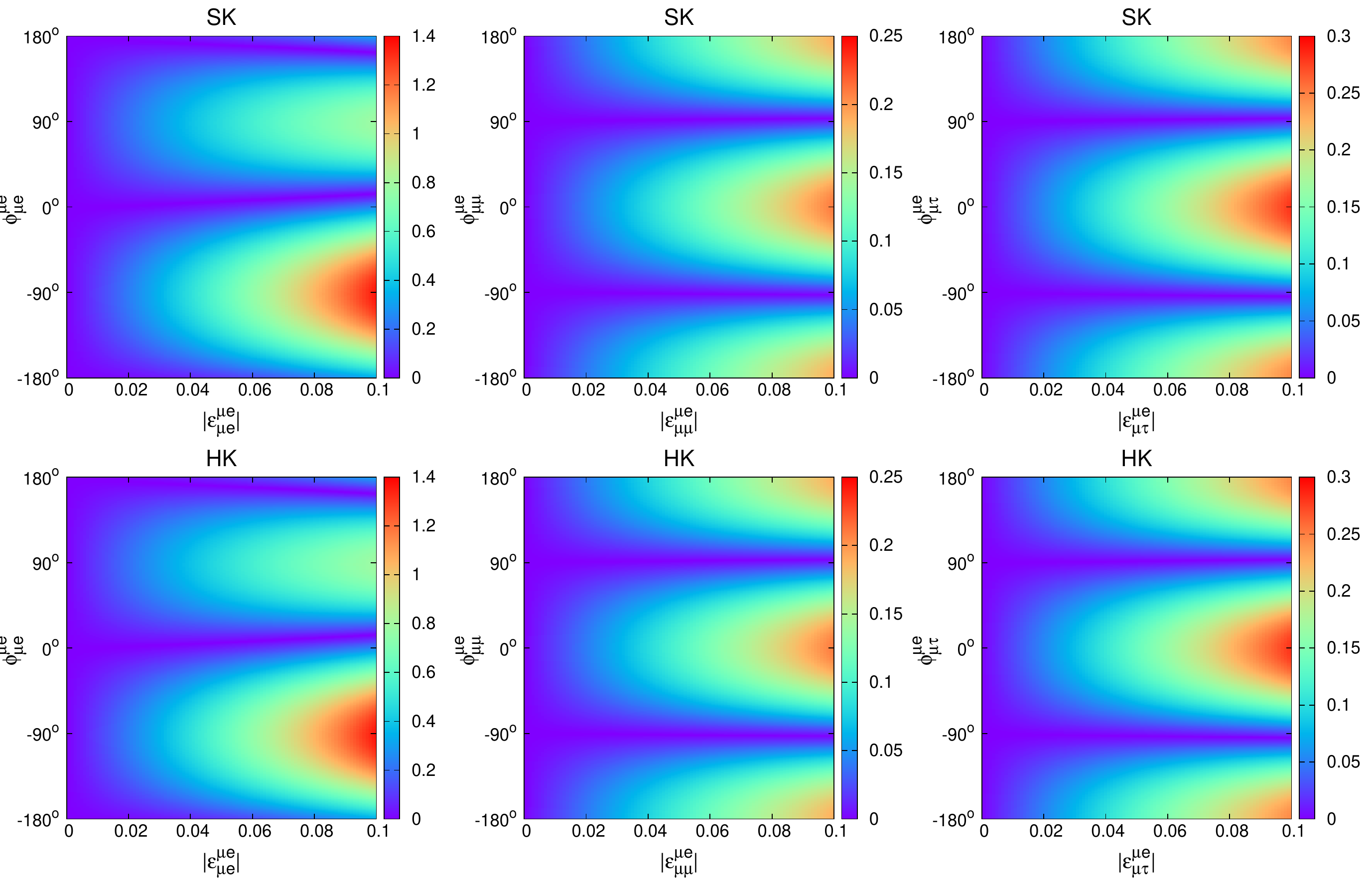}\end{center}
\caption{\footnotesize {Oscillograms of ${|P_{NSI}-P_{SI}|}/{P_{SI}}$ at  SK (HK) with corresponding baseline of 15 (23) km and energy of 35 (45) MeV in the upper (bottom) row. The three columns depict the effect of source NSI parameters $\epsme$, $\epsmm$ and $\epsmt$. Note the difference in scale among the rows.}}
\label{fig1}
\end{figure*}

\begin{figure}[!htb]
\begin{center}
\includegraphics[width=5.4cm,height=5.5cm]{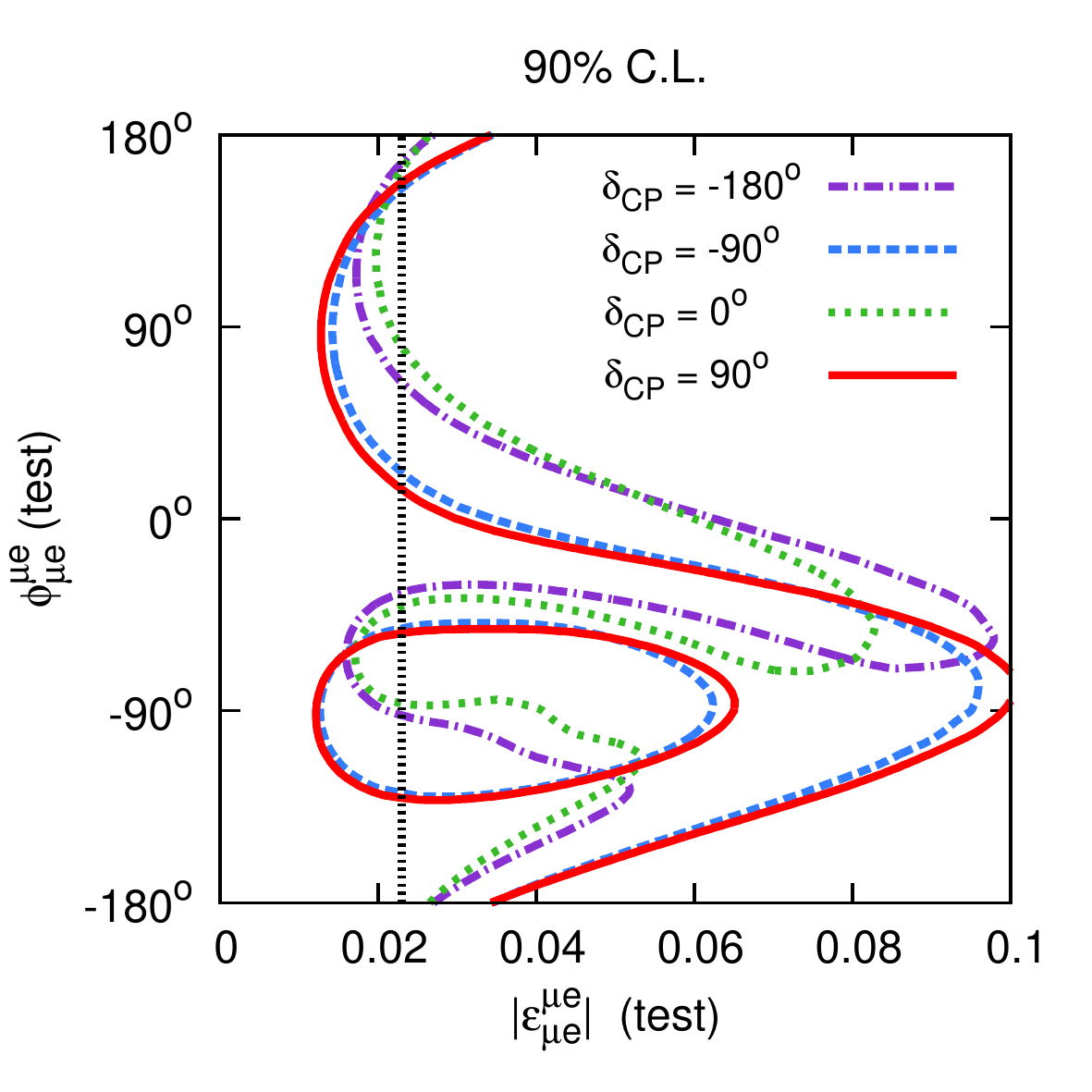}
\includegraphics[width=5.4cm,height=5.5cm]{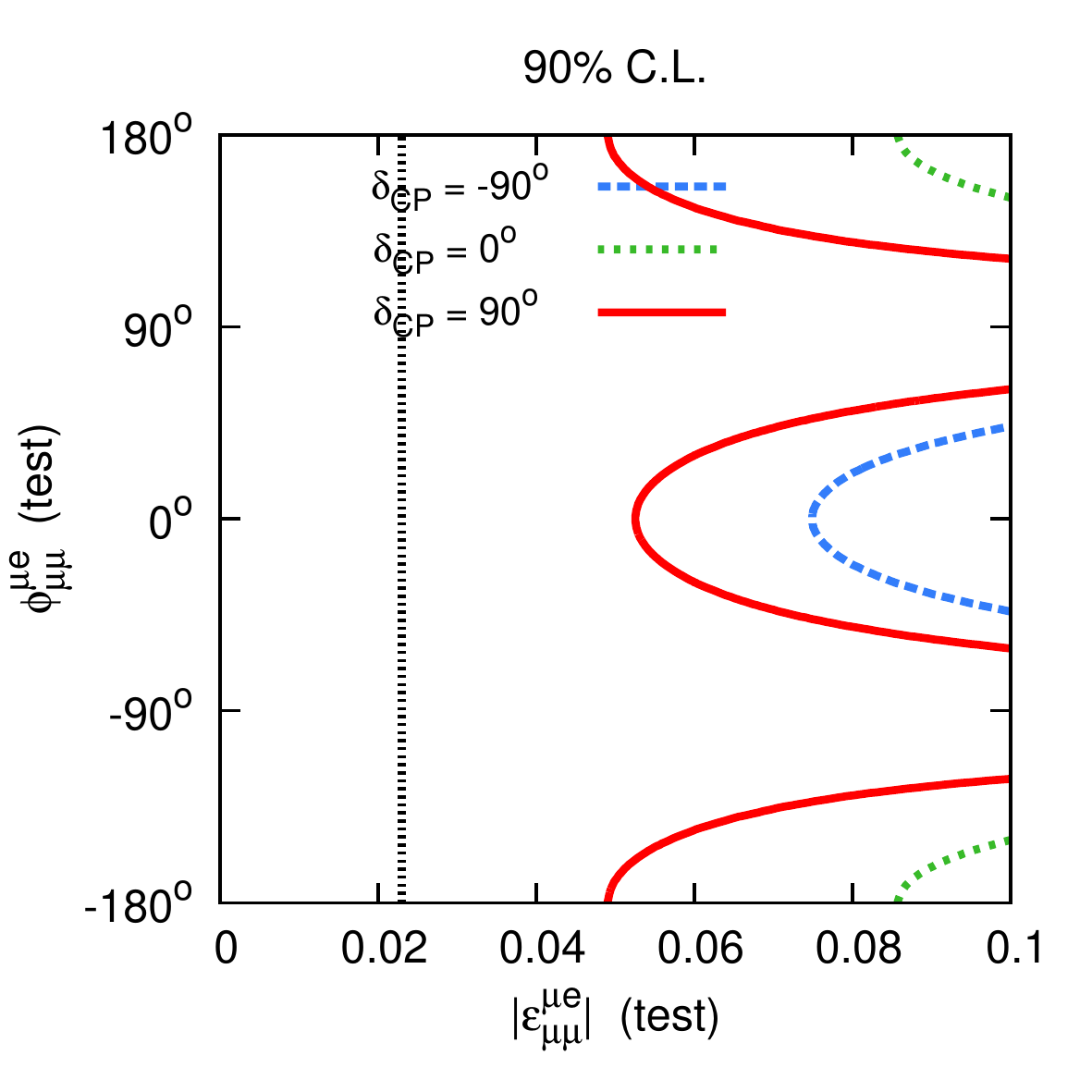}
\includegraphics[width=5.4cm,height=5.5cm]{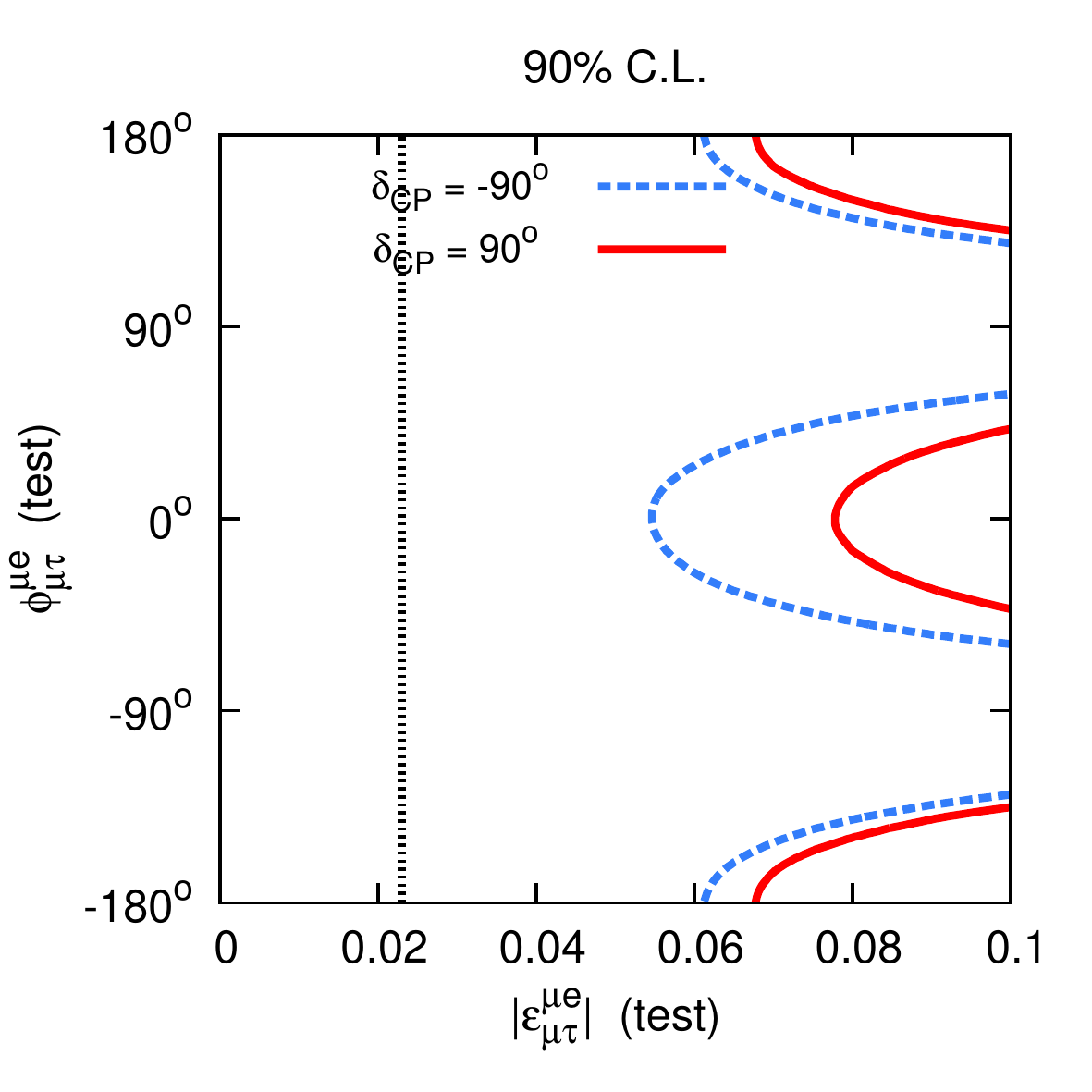} \\
\includegraphics[width=5.4cm,height=5.5cm]{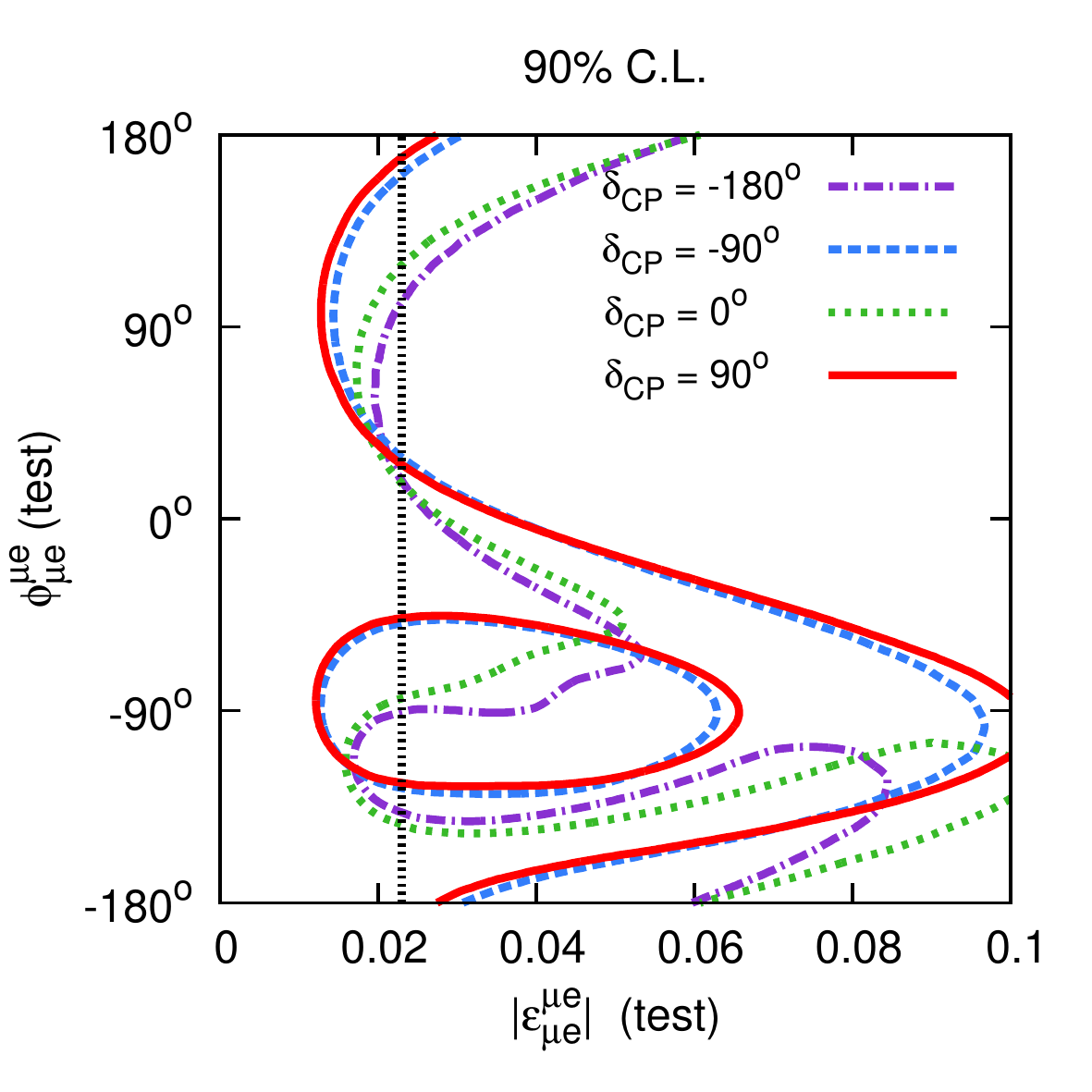}
\includegraphics[width=5.4cm,height=5.5cm]{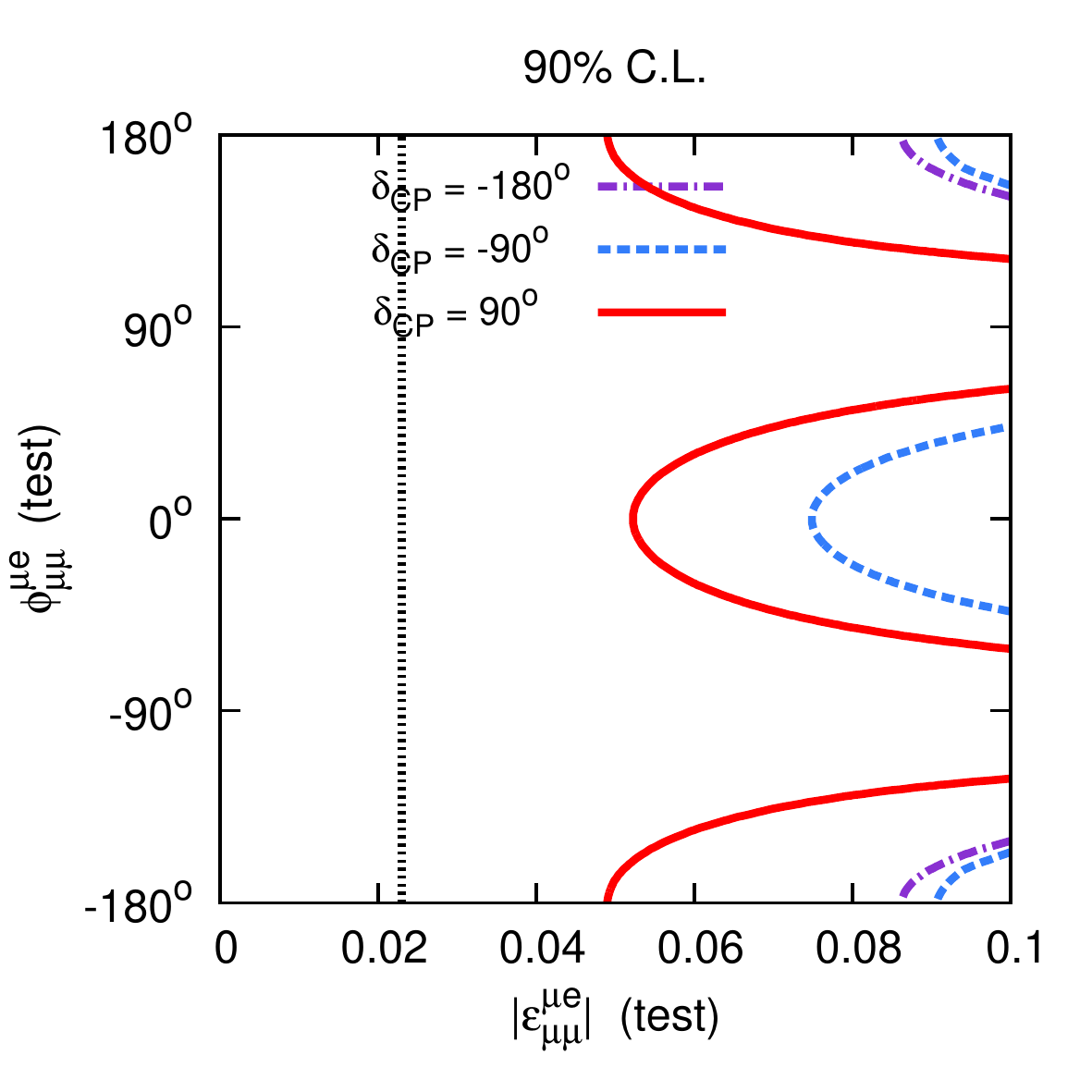}
\includegraphics[width=5.4cm,height=5.5cm]{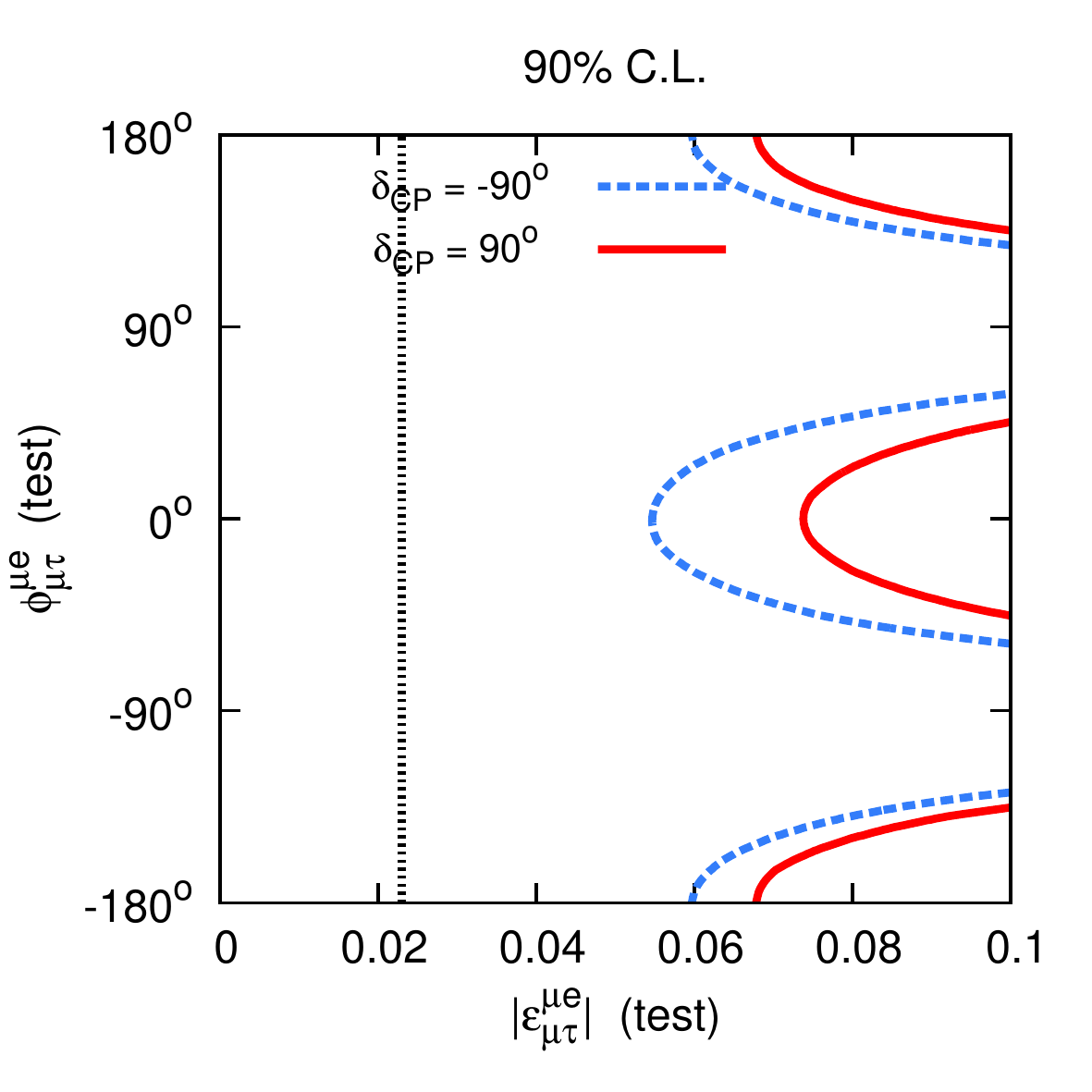} \\
\end{center}
\caption{\footnotesize The bounds on source NSIs at  \mudar\  facility (SK and HK) in  $\mu-e$, $\mu- \mu$, and $\mu-\tau$ sectors are  given in left, middle and right panels respectively. The neutrino mass hierarchy is assumed to be NH (IH) in the upper (lower) panel.}
\label{fig:newPnue}
\end{figure}
In the present study, we study the impact of NSIs in the context of a \mudar\  facility,  which was conceived for clean and precise measurement of the parameters (especially $\dcp$) in standard neutrino oscillation scenario~\cite{Alonso:2010fs,Evslin:2015pya,Agarwalla:2010nn}. In particular, we   address the following  questions which are relevant -
\begin{enumerate}
\item[(1)]  How well we can measure or constrain the relevant NSI parameters at such a facility ? 
\item[(2)]  What is the impact of NSIs on the measurement of the standard CP phase, $\dcp$ ?
\end{enumerate}

Let us begin with a  discussion  of (1) which is about the constraints that can be placed on  the relevant NSI parameters at a \mudar\ facility. We end this section with a discussion of the second question  in (2) which pertains to testing the robustness of the precision with which $\dcp$ can be measured at such a facility.

The numerical computation is carried out using the  General Long Baseline Experiment Simulator  (GLoBES)~\cite{Huber:2004ka} package along with the new physics plugin for sterile neutrinos and NSIs~\cite{Kopp:2006wp}. The neutrino oscillation parameters  used in our analysis are given in Table~\ref{tab:parameters}.\footnote{We have checked that  the oscillation parameters that  significantly impact the  results  are $\theta_{23}$ and $\dcp$.  Marginalization over $\dm{31}$ does not affect the results much and therefore we keep it fixed.}

In order to comprehend our results better, let us first attempt to understand the  expected sensitivity at the probability level to the NSI parameters at the \mudar\ setup. In Fig.~\ref{fig1},  we plot in the plane of $|\varepsilon^{\mu e}_{\mu \beta}| - \phi^{\mu e}_{\mu \beta}$, the relative absolute difference of the probabilities 
 $({| P_{SI} - P_{NSI} |}/{P_{SI}})$   where  $P_{SI}$ ($P_{NSI}$) and the oscillation probability with SI  (NSI) effects.   
 The top (bottom) row depicts the effect at   SK (HK) which corresponds to a baseline of 15 (23) km and energy of 35 (45) MeV.
   The three columns correspond to the different source NSI parameters, $\epsme$, $\epsmm$ and $\epsmt$ respectively. 

The first  observation is that the \mudar\ is most sensitive to the source NSI parameter, $|\epsme|$. The sinusoidal nature of the curve in the first column can be easily understood from the $\sin\phime$ dependence in Eq.~\ref{eq:pme} at the oscillation maximum (when $\Delta=\pi/2$) and for the currently favored value of $\dcp=-90^\circ$. From Eq.~\ref{eq:pmeformal}, we see that the probability depends only on the real part of $\epsmm$; hence, the second column shows a $\cos\phimm$ dependence, with maximum at $\phimm=0,180^\circ$. This same feature is also seen for $\epsmt$ in the last column. We will see later that these characteristic shapes are also reflected in the $\chi^2$ analysis.

Next, we discuss the bounds on the NSI parameters. We present our results for both normal hierarchy (NH) and inverted hierarchy (IH), and assume in each case that the hierarchy is known from other experiments, i.e., we do not marginalize over  hierarchy. To obtain a bound on any of the NSI parameters using \mudar\ facility and either SK or HK, we generate the true event spectrum with the standard oscillation framework using the parameters listed in Table~\ref{tab:parameters} and compare it with a test event spectrum which is simulated with NSI scenario  with only one of the NSI parameters taken as nonzero at a time. 
  We show the 90\% C.L. exclusion regions in the plane of $|\varepsilon_{\alpha\beta}^{\mu e}|-\phi_{\alpha\beta}^{\mu e}$  for each of the three sectors ($\mu - e $, $\mu- \mu$ and $\mu -\tau$ respectively) as shown in Fig.~\ref{fig:newPnue}. 
  The total $\chi^2$ is obtained by adding the individual contributions from \mudar\ at SK and  \mudar\ at HK    for each point in the test parameter space,
\begin{equation}
\chi^2_{total}=\chi^2_{\textrm{\mudar@}SK}+ \chi^2_{\textrm{\mudar@}HK} \nonumber
\end{equation} 
We add a Gaussian prior for $\sin^2\theta_{23}$ with $\sigma(\sin^2\theta_{23}=0.015)$ and marginalize over $\dcp$ and $\sin^2\theta_{23}$ in their allowed range in Table~\ref{tab:parameters}.\footnote{Since at a \mudar\ facility,  it is not possible to study $\nu_\mu \to \nu_\mu$ oscillations, we have added a prior on  $\sin^2\theta_{23}$.}
  The bounds are presented  for four different true values of $\dcp$ i.e., $-180^\circ$, $-90^\circ$, $0^\circ$, and $90^\circ$. The left, middle and the right panels correspond to the NSI parameters $\epsme$, $\epsmm$, and $\epsmt$, respectively. 
   In all the three panels, the black dotted vertical line indicates the present 90\% C.L. upper bound of the parameter. 

\begin{figure}[!htb]
\begin{center}
\hspace{-0.5 in}
\includegraphics[width=5.5cm,height=5.5cm]{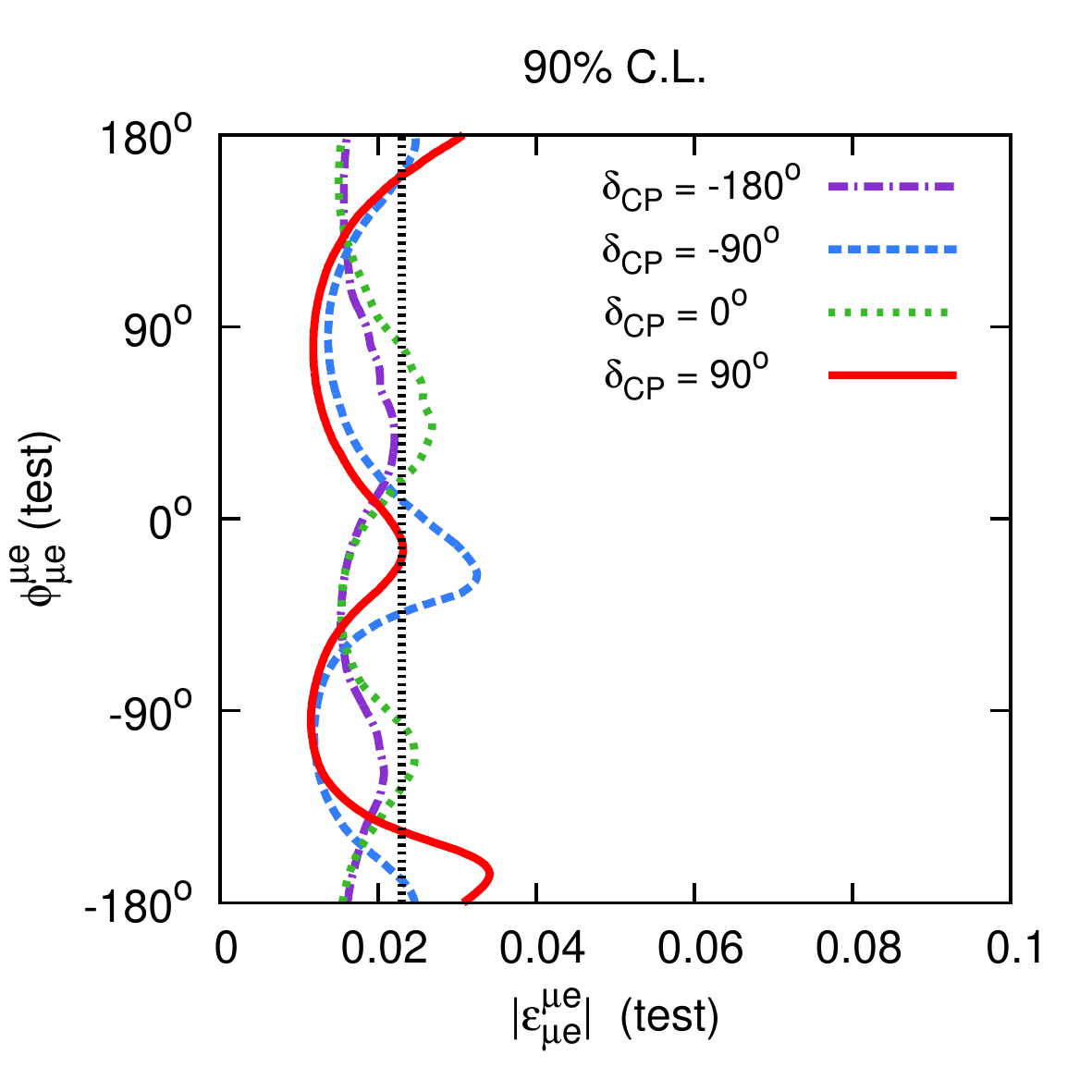}
\includegraphics[width=5.5cm,height=5.5cm]{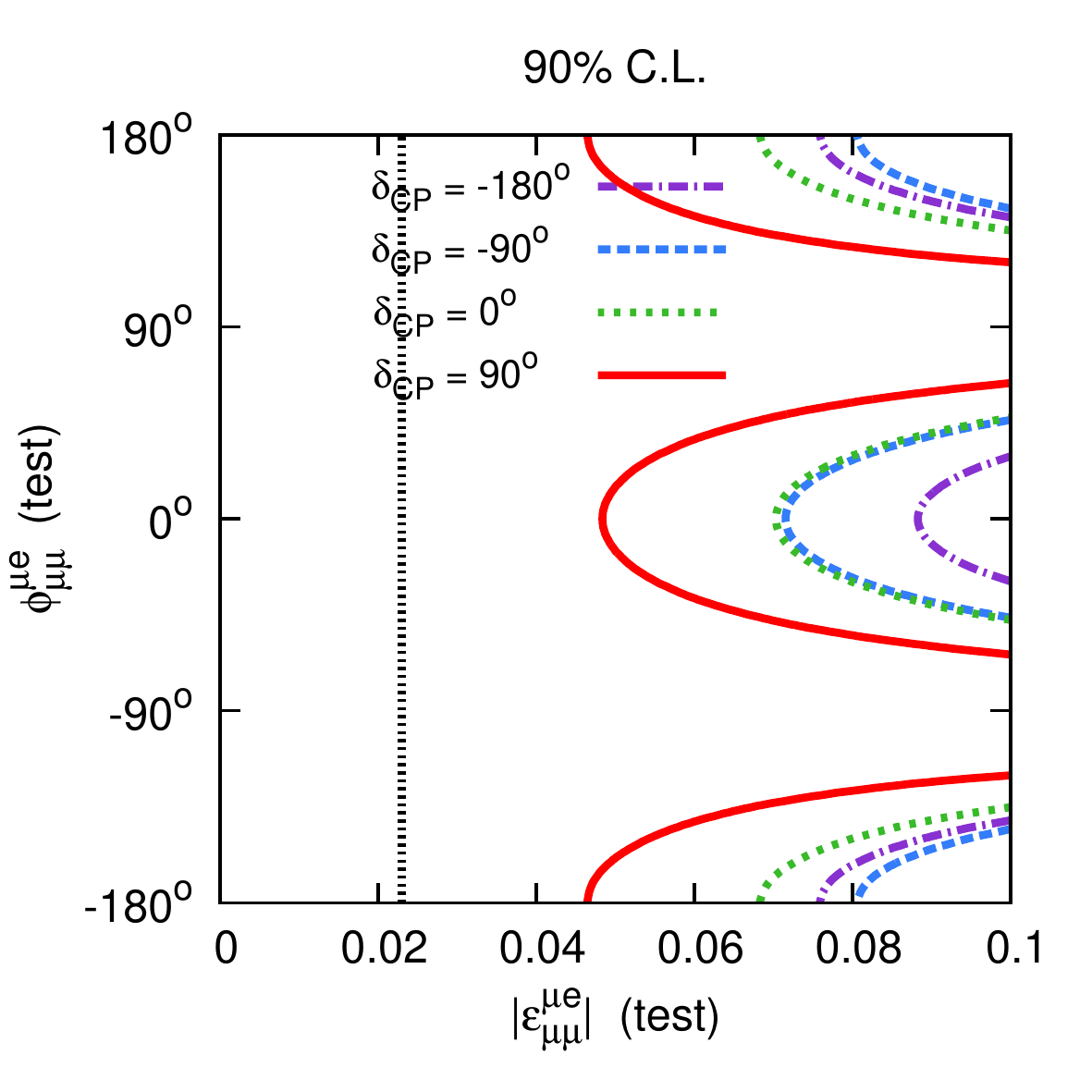}
\includegraphics[width=5.5cm,height=5.5cm]{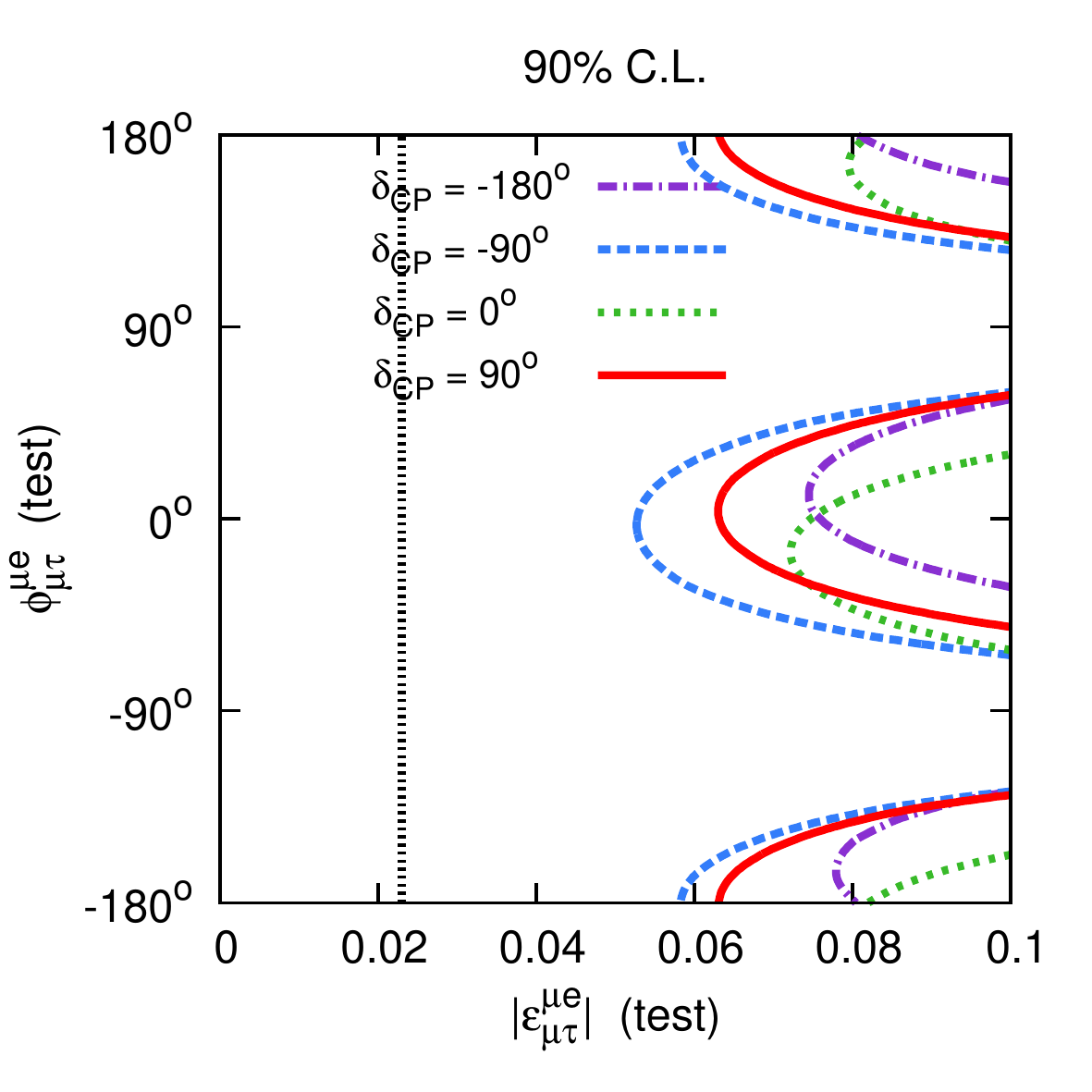} \\
\hspace{-0.5 in}
\includegraphics[width=5.5cm,height=5.5cm]{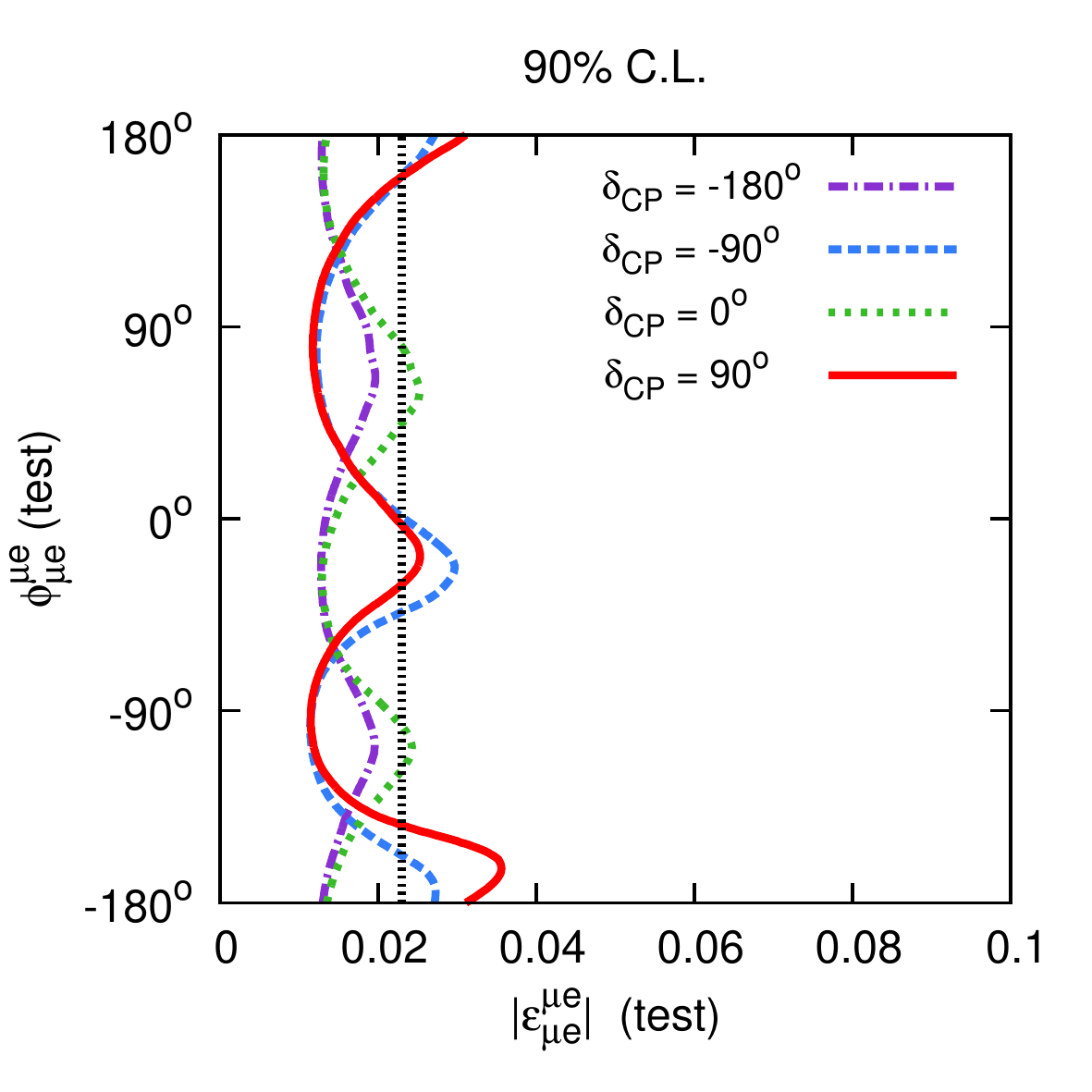}
\includegraphics[width=5.5cm,height=5.5cm]{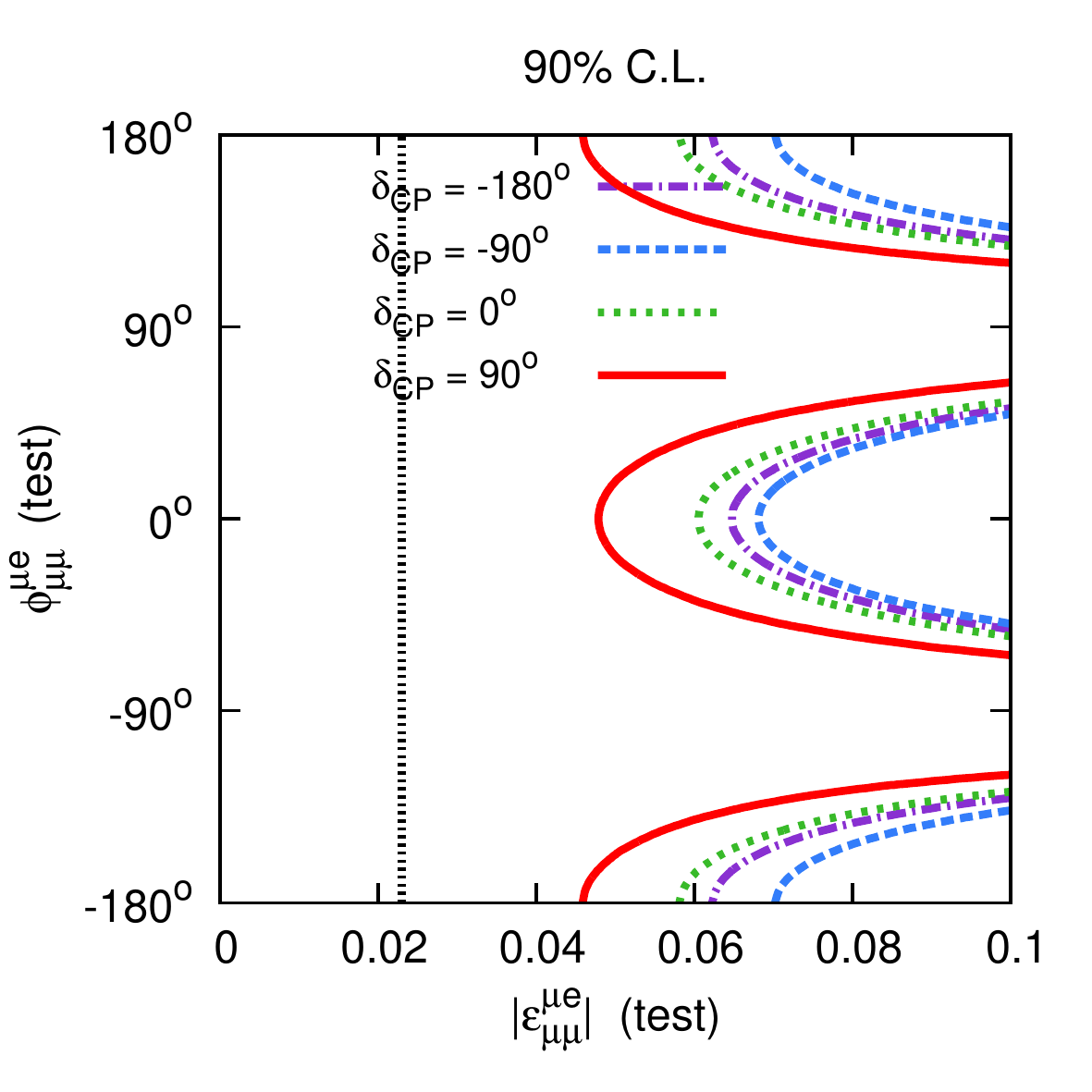}
\includegraphics[width=5.5cm,height=5.5cm]{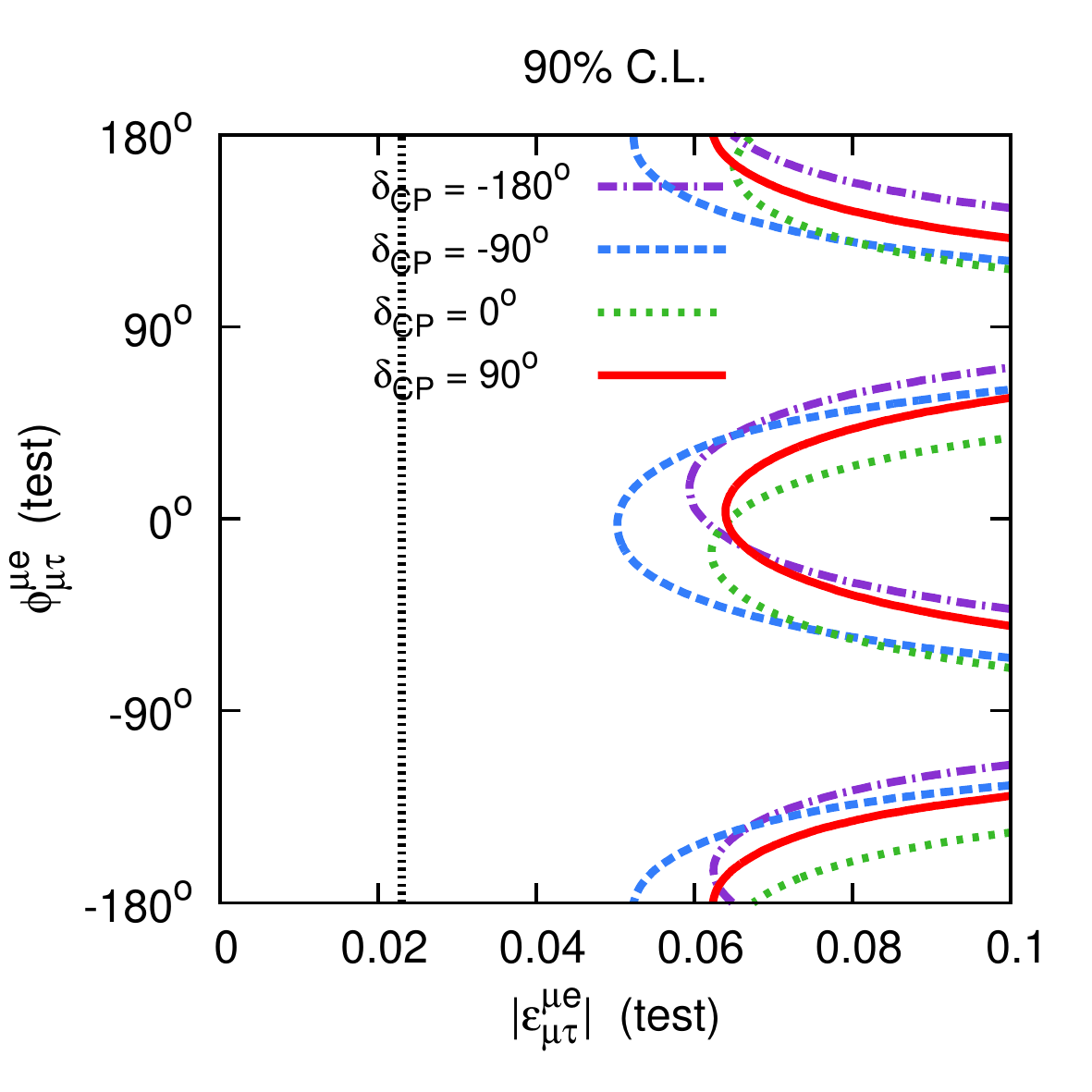} \\
\end{center}
\caption{\footnotesize The bounds on source NSIs at T2HK + \mudar\ (SK and HK) in  $\mu-e$, $\mu- \mu$, and $\mu-\tau$ sectors are   given in left, middle and right panels respectively. Top row corresponds to T2HK running fully in neutrino mode while the bottom row corresponds to when the neutrino to antineutrino running ratio of T2HK is 1:3. }
\label{fig:combined}
\end{figure}

   From Fig.~\ref{fig:newPnue}, we can make the following observations. The value of the NSI phase plays an important role as the sensitivity to NSI in a given sector is very much dependent on the value of the NSI phase. 
\begin{itemize}
\item $\mu -e$ sector :
The left panel depicts the bounds on $|\epsme|$ as a function of $\phime$. Marginalization over the unknown test value of $\dcp$ results in degeneracy between $\dcp$ and $\phime$ at higher values of $|\epsme|$, resulting in the features seen in the plot. Consequently, values of $\phime$ in the upper half-plane are more constraining for this sector, especially for $\dcp=\pm90^\circ$. We obtain a 90\% C.L. exclusion for $|\epsme| < 0.012$  for $\dcp=\pm 90^\circ$ if the associated value of $\phime$ is $+90^\circ$ for \mudar. However, in the absence of any information about $\phime$, we obtain an upper bound of $|\epsme| < 0.1$.

%We obtain an upper bound, $|\epsme| < 0.012$ for $\dcp=\pm 90^\circ$. This is clearly an  improvement over the existing bounds.

It is worth mentioning  that if the value of $\dcp$ is assumed to be known and fixed at its true value, the bound oscillates mildly {{between 0.01 and 0.02}} for all values of $\phime$. A similar effect will be seen later in   Fig.~\ref{fig:combined}, where addition of T2HK data leads to shrinking of contours due to  lifting of the parameter degeneracies, thereby leading to results that are fixed parameter like.

\item $\mu -\mu$ sector : From the middle   panel, we note that for $\epsmm$ the sensitivity is better at CP-conserving values, $\phimm = 0^\circ$, $180^\circ$, but the bounds obtained from \mudar\ facility are weaker than the existing ones.

\item $\mu -\tau$ sector : The conclusions are similar to those for the $\mu - \mu $ sector as mentioned above.

\item  The bounds for IH as seen in the lower panels are qualitatively similar to the ones for NH, with variations depending  on the choice of $\dcp$ and NSI phase. 

\end{itemize}

The peculiar shape of the contour seen in the left panel is due to the parameter degeneracy between $|\epsme|$, $\phime$ and $\dcp$, and can be explained with the help of Eq.~\ref{eq:pme}. For the sake of simplicity, we assume the energy to be fixed at the oscillation maximum ($\Delta=\pi/2$). The formula reduces to 
\begin{eqnarray}
 \overline{P}^{NSI} & = & 4 \sin^2\theta_{13} \sin^2\theta_{23} + \pi \alpha\sin\theta_{13} \sin 2\theta_{12} \sin 2\theta_{23} \sin\dcp \nonumber \\
 & - & 4 |\epsme| \sin\theta_{13} \sin\theta_{23} \cos(\dcp+\phime) + \pi |\epsme| \alpha \sin 2\theta_{12} \cos\theta_{23} \sin\phime \nonumber \\
 & \equiv & P_0 + P_1 \sin\dcp - Q_1 |\epsme| \cos(\dcp+\phime) + Q_2 |\epsme| \sin\phime ~,
\end{eqnarray}
where $P_0$, $P_1$, $Q_1$ and $Q_2$ are strictly positive.
As usual, the standard probability is recovered by setting $|\epsme| \to 0$. Since $Q_2$ is quite small, 
the second and third terms dictate the interplay among the parameters.

The standard probability when $\dcp=\pm90^\circ$ is simply $P_0 \pm P_1$, which can be made degenerate with the NSI probability $P_0 + P_1 \sin\dcp - Q_1 |\epsme| \cos(\dcp+\phime)$ by a suitable choice of $\dcp$ and $\phime$. However, this only works when $|\epsme|$ is large enough to compensate for the difference between the coefficients $P_1$ and $Q_1$. Therefore, we see a degenerate band around $|\epsme|=0.08$ and $\phime=-90^\circ$. The value $|\epsme|=0.04$ is not large enough for this effect to be seen and hence is excluded. 

The results of the analysis can be improved by constraining the values of the other oscillation parameters (in particular, $\dcp$) that affect the probability in the multidimensional parameter space. It should be noted that during the running of \mudar\ facility, the T2HK experiment will be also taking data. Therefore, we discuss the possibility of adding the T2HK data to the \mudar\ data to improve the sensitivity to considered NSI parameters. At T2HK, neutrinos are produced from pion decay and are hence unaffected by the $\varepsilon^{\mu e}_{\alpha\beta}$ parameters which are specific to muon-decay production mechanism.
However, its ability to discriminate between true and spurious degenerate regions in the standard oscillation parameter space leads to a synergy with \mudar\ facility and can improve the sensitivity to the NSI parameters. In the case of T2HK experiment, we simulate both true and test event spectra without new physics and obtain the $\chi^2$ by comparing these event spectra. The total $\chi^2$ is obtained by adding the individual contributions from \mudar\ at SK, \mudar\ at HK  and T2HK for each point in the test parameter space,
\begin{equation}
\chi^2_{total}=\chi^2_{\textrm{\mudar@}SK}+ \chi^2_{\textrm{\mudar@}HK}+ \chi^2_{T2HK} \nonumber
\end{equation} 

\begin{figure}[!htb]
\begin{center}
 \includegraphics[width=0.95\textwidth]{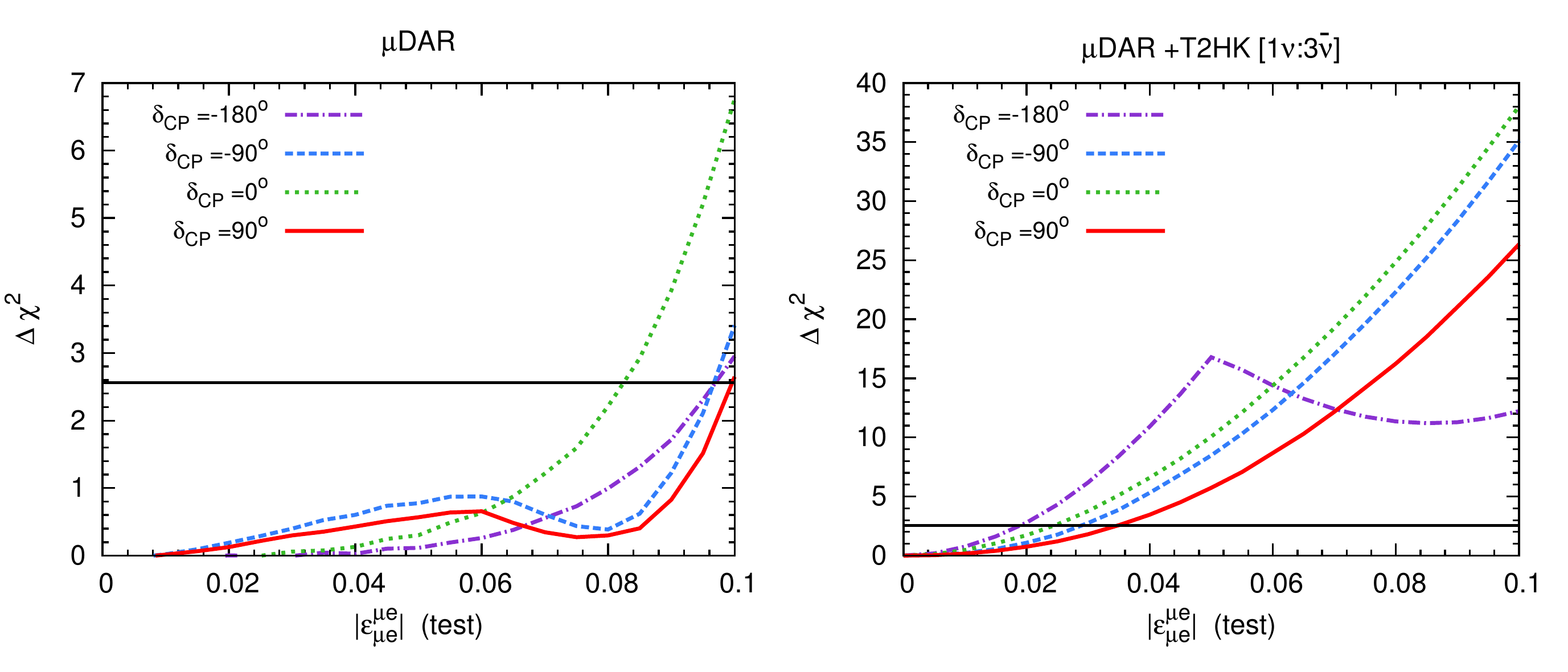}
% \includegraphics[width=0.45\textwidth]{
 %T2HK-DAR_mu_e-ME}
 % T2HK_DAR_mu_e-ME.pdf }
\end{center}
\caption{\footnotesize Bounds on $|\epsme|$ in absence of any information about the associated phase, $\phime$. The various curves represent different values of $\dcp$. The left panel is for \mudar\, alone while the right panel is for \mudar\, + T2HK with the runtime   split into neutrino mode and antineutrino mode as depicted in the plot.}
\label{NSI_bound}
\end{figure}

We present our results in Fig.~\ref{fig:combined}. In the top row, we have given the results with T2HK running fully in the neutrino mode with the addition of \mudar\  and in the bottom row, we have given the results when the neutrino to antineutrino runtime ratio at T2HK is 1:3 with the addition of \mudar. In each row the left, middle, and right panels correspond to NSI parameters $\epsme$, $\epsmm$, and $\epsmt$, respectively. From the figure, we observe that the addition of T2HK data significantly improves the bounds on the NSI parameter $|\epsme|$. This is achieved due to the synergy between the \mudar\ facility and T2HK experiment in constraining $\dcp$. The resulting bounds lie between 0.01 and 0.02 depending on the values of the two phases, which is better than the existing bound over most of the parameter space. There is also an improvement in the bounds on $|\epsmm|$ and $|\epsmt|$, significantly for values of $\dcp$ in the lower half-plane. However, for these two parameters the bounds do not improve the 
existing ones. This is expected from the fact that these play a subleading role, as discussed in Sec.~\ref{sec:formalism}. The results for $\epsmm$ and $\epsmt$ can be improved slightly by changing the neutrino to antineutrino ratio, as seen in the lower panels. {The results for T2HK neutrino to antineutrino ratio of 1:3 is better than T2HK pure neutrino run because of the fact that the antineutrino run helps to resolve the octant degeneracy and hence improving the $\dcp$ measurement \cite{Ghosh:2014zea}.}

Figures~\ref{fig:newPnue} and~\ref{fig:combined} depict the bounds on the NSI parameters for a certain choice of the associated NSI phase. In order to obtain the bounds on the NSI parameters in the absence of any information about the NSI phases, in Fig. \ref{NSI_bound}, we have presented the $\chi^2$ vs $\epsme$ plot by marginalizing over $\phime$ for the case of NH. The left panel is for \mudar\ and the right panel is for T2HK + \mudar\ . From the two panels, we observe that   for  \mudar\ alone,  the upper bound obtained for $\epsme$ at 90\% C.L is weaker than the present bound. But, in a combination of T2HK with \mudar\,  the upper bound of  $\epsme$ at 90\% C.L is somewhat stronger than the present bound. The best upper bound obtained is  $|\epsme| < 0.02$ for $\dcp = -180^\circ$.

%-------------------------------------------
 
\begin{figure}[!htb]
\begin{center}
 \includegraphics[width=0.45\textwidth]{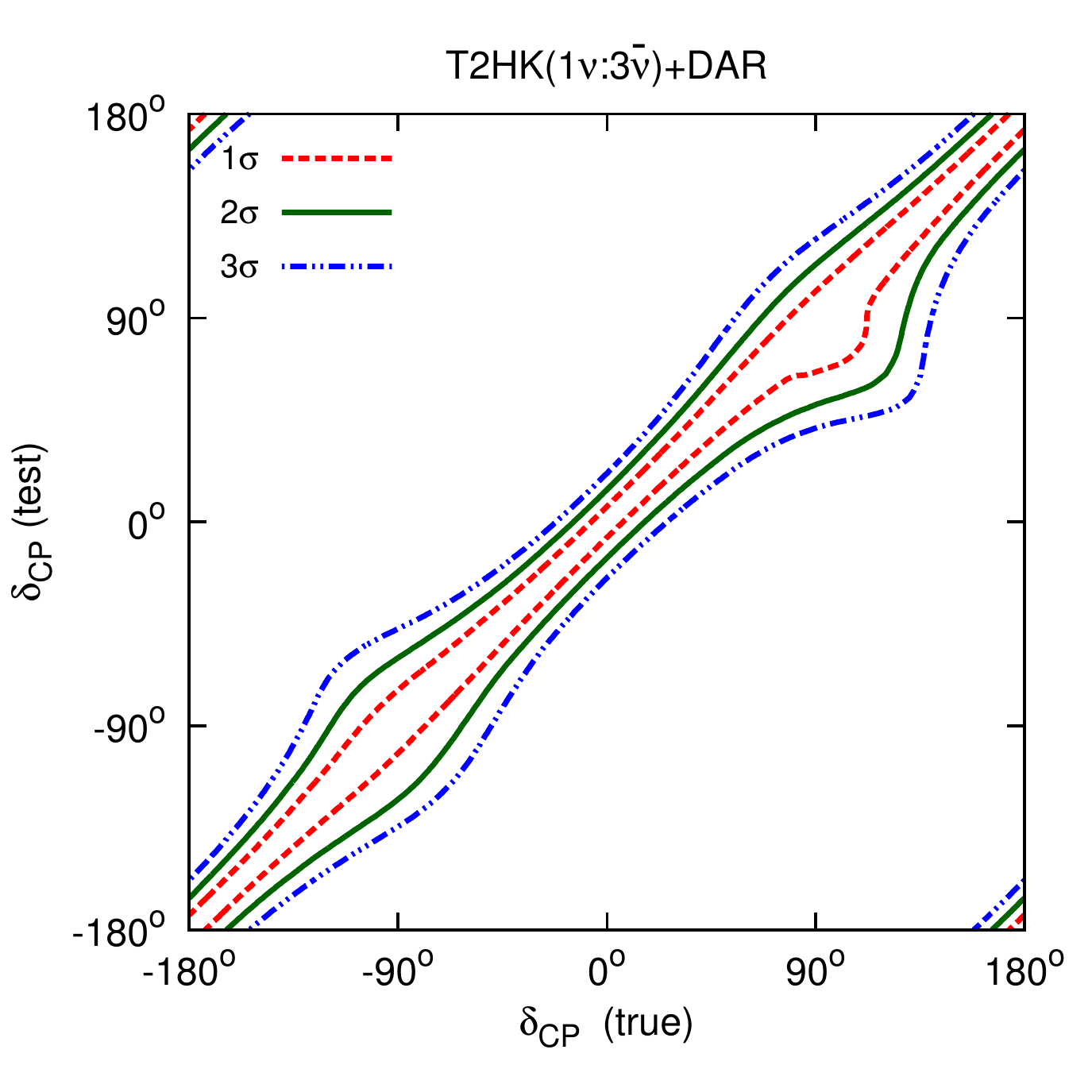}
 \includegraphics[width=0.45\textwidth]{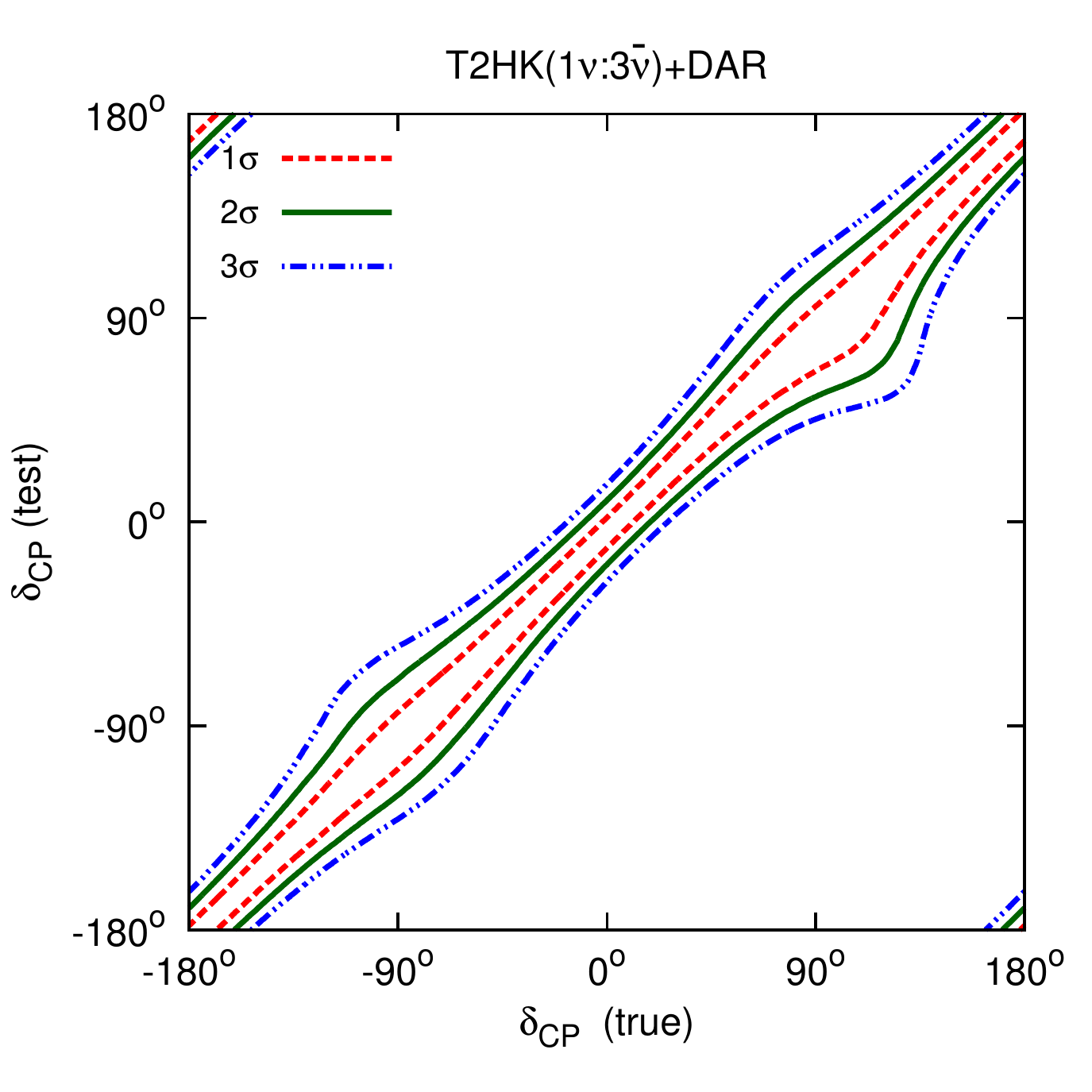}
\end{center}
\caption{\footnotesize Range of allowed values of $\dcp$ as a function of its true value. The true values of NSI parameters are $\epsme=0.01$ and $\phime=0$. Left panel: fit against oscillations in the presence of NSI; right panel: fit against standard oscillations.}
\label{fig:dcpprec}
\end{figure}

We now address the question pertaining to the robustness of  $\dcp$ precision, i.e., what is the impact of  inclusion of additional NSI parameters on the measurement of standard CP violation at the combined setup T2HK(1:3)+\mudar\ facility. The range of allowed values of $\dcp$ is shown as a function of its true value in Fig.~\ref{fig:dcpprec}. The true values of NSI parameters are chosen to be $\epsme=0.01$ and $\phime=0$ for illustrative purposes. From the left panel, we see the result of fitting the simulated data against the test hypothesis including the NSI. The allowed values of $\dcp$ are very closely correlated with its true value, signaling the robustness of the $\dcp$ measurement at this setup. However, when data from this setup are first analyzed, it is likely to be fitted against standard oscillations without NSIs. The result of such a fit is shown in the right panel. Once again, we find allowed regions around the true value of $\dcp$. Any strong degenerate effects in the parameter space would 
show up as allowed regions away from the $y=x$ line in these panels. Their absence indicates the robustness of $\dcp$ measurement. Similar results can be generated for various chosen values of $\phime$. In Table~\ref{tab:prec}, we list the precision in $\dcp$ for various choices of the true values of $\dcp$ and $\phime$.

\begin{table*}[htb]
 \begin{tabular}{|c|c|c|c|c|}
  \hline
  \multirow{3}{*}{True $\dcp$ [${}^\circ$]} &  \multirow{3}{*}{True $\phime$ [${}^\circ$]} & \multicolumn{3}{c|}{$\dcp$-precision [${}^\circ$]} \\
  \cline{3-5}
  && Standard oscillations & \multicolumn{2}{c|}{NSI scenario} \\
  \cline{3-5}
  && T2HK+\mudar & \mudar & T2HK+\mudar \\
  \hline
  \multirow{4}{*}{-90} & -90 & \multirow{4}{*}{57} & 103 & 65 \\
  \cline{2-2} \cline{4-5}
  & 0 & & 77 & 57 \\
  \cline{2-2} \cline{4-5}
  & 90 & & 59 & 49 \\
  \cline{2-2} \cline{4-5}
  & 180 & & 65 & 52 \\
  \hline
  \multirow{4}{*}{0} & -90 & \multirow{4}{*}{30} & 88 & 19 \\
  \cline{2-2} \cline{4-5}
  & 0 & & 60 & 30 \\
  \cline{2-2} \cline{4-5}
  & 90 & & 46 & 27 \\
  \cline{2-2} \cline{4-5}
  & 180 & & 59 & 30 \\
  \hline
  \multirow{4}{*}{90} & -90 & \multirow{4}{*}{55} & 108 & 61 \\
  \cline{2-2} \cline{4-5}
  & 0 & & 83 & 55 \\
  \cline{2-2} \cline{4-5}
  & 90 & & 66 & 50 \\
  \cline{2-2} \cline{4-5}
  & 180 & & 71 & 52 \\
  \hline
  \multirow{4}{*}{180} & -90 & \multirow{4}{*}{30} & 92 & 32 \\
  \cline{2-2} \cline{4-5}
  & 0 & & 65 & 30 \\
  \cline{2-2} \cline{4-5}
  & 90 & & 47 & 27 \\
  \cline{2-2} \cline{4-5}
  & 180 & & 59 & 30 \\
  \hline
 \end{tabular}
\caption{90\% C.L. $\dcp$ precision for various true values of phases $\dcp$ and $\phime$.}
\label{tab:prec}
\end{table*}

\begin{figure}[!htb]
\begin{center}
 \includegraphics[width=0.4\textwidth]{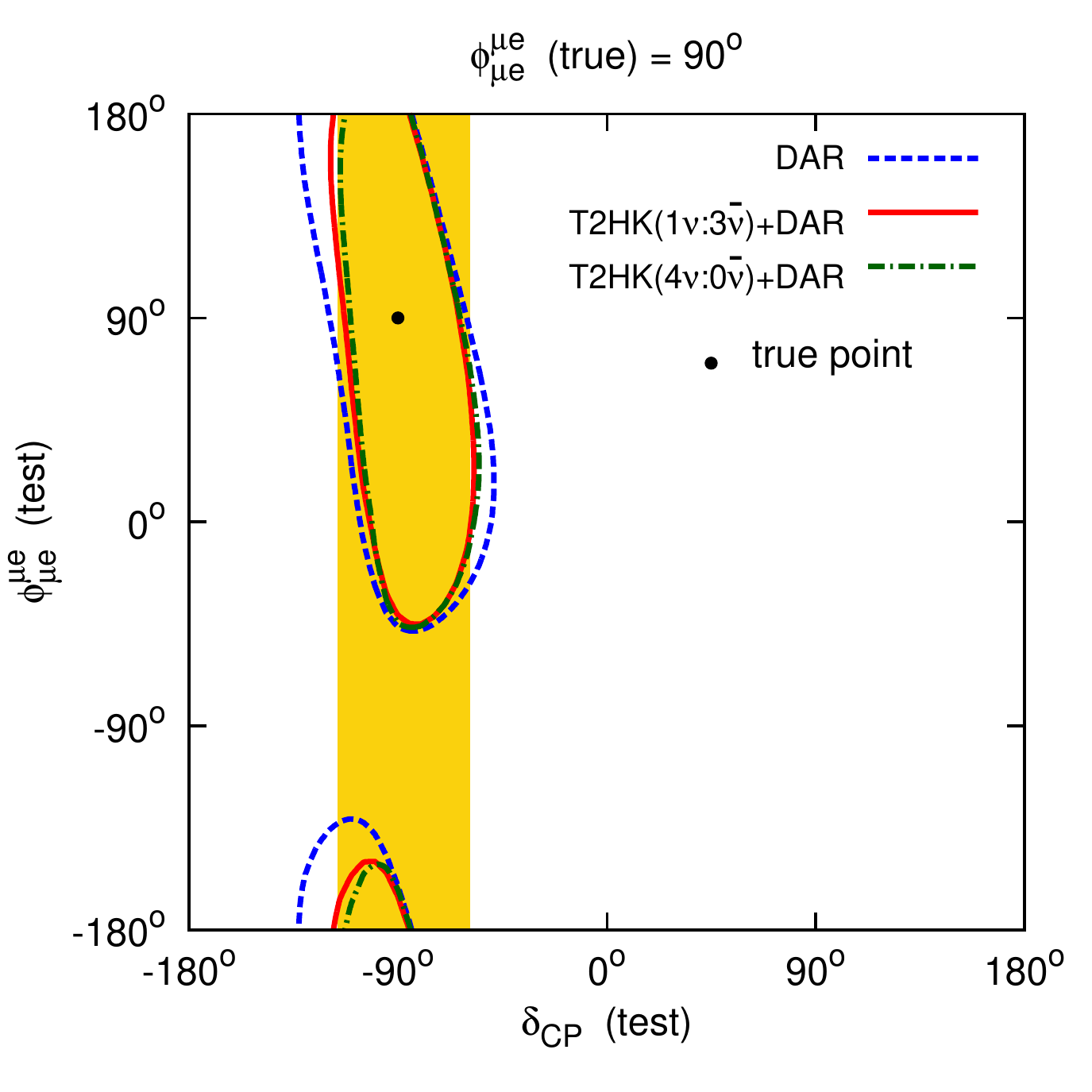} 
 \includegraphics[width=0.4\textwidth]{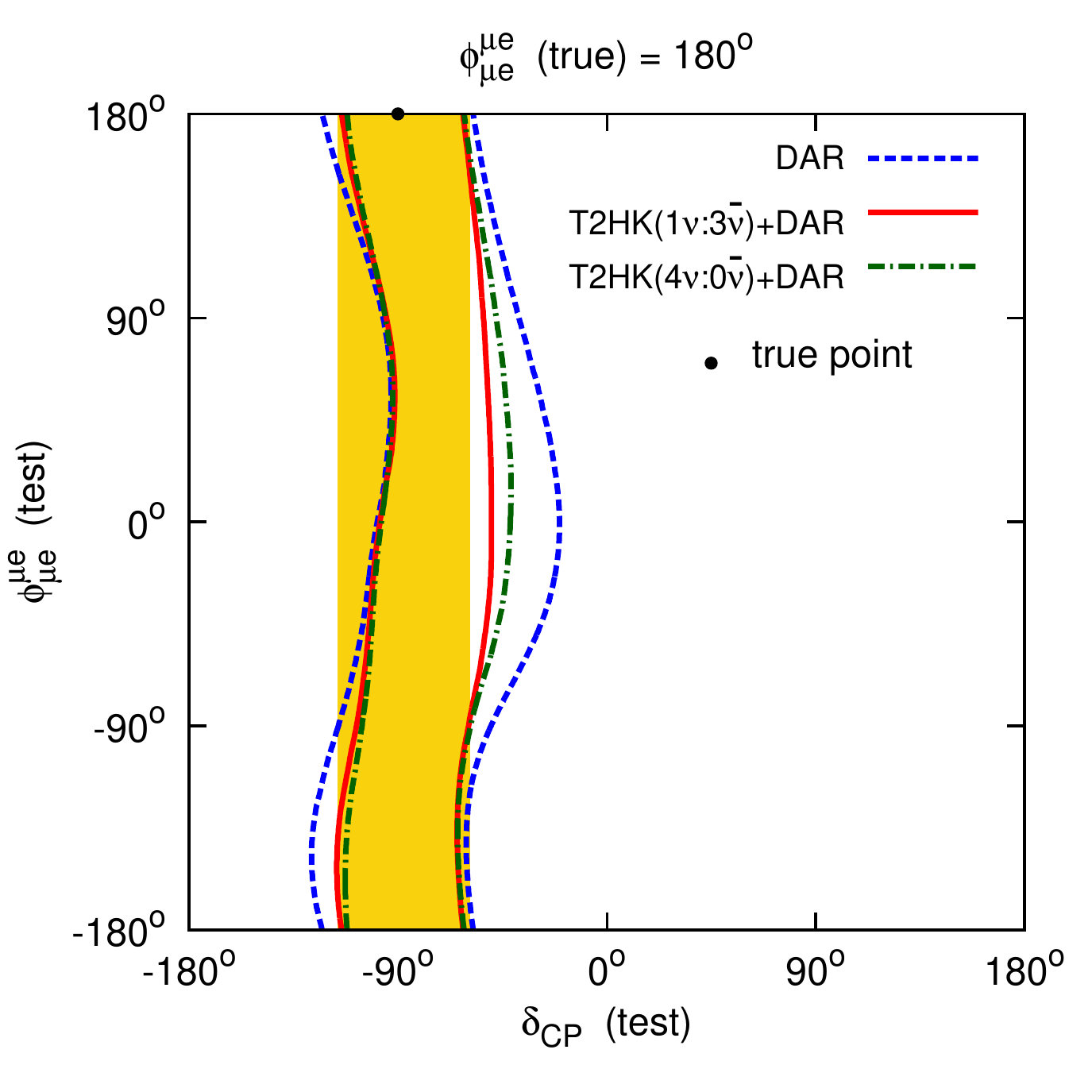}\\
 \includegraphics[width=0.4\textwidth]{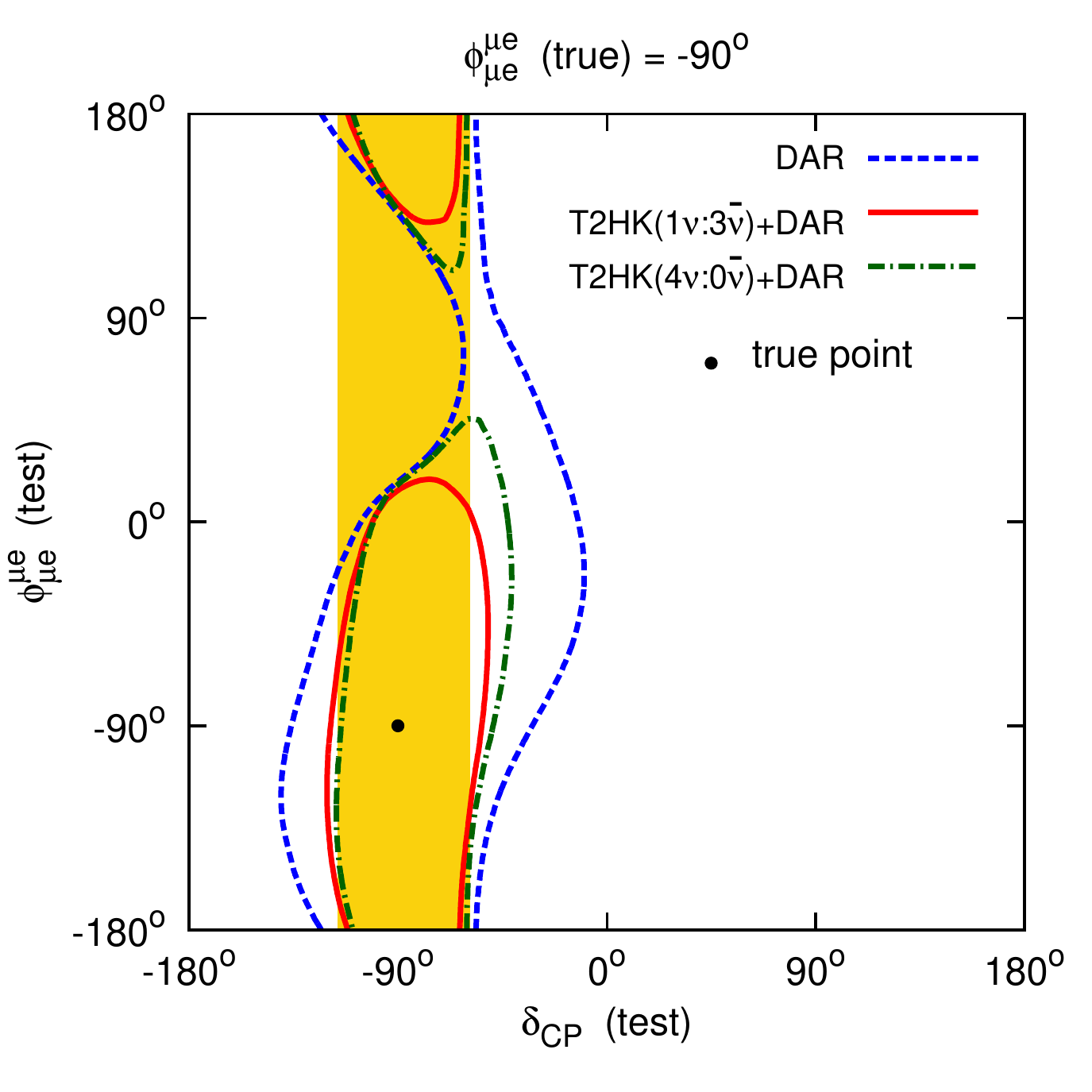}
 \includegraphics[width=0.4\textwidth]{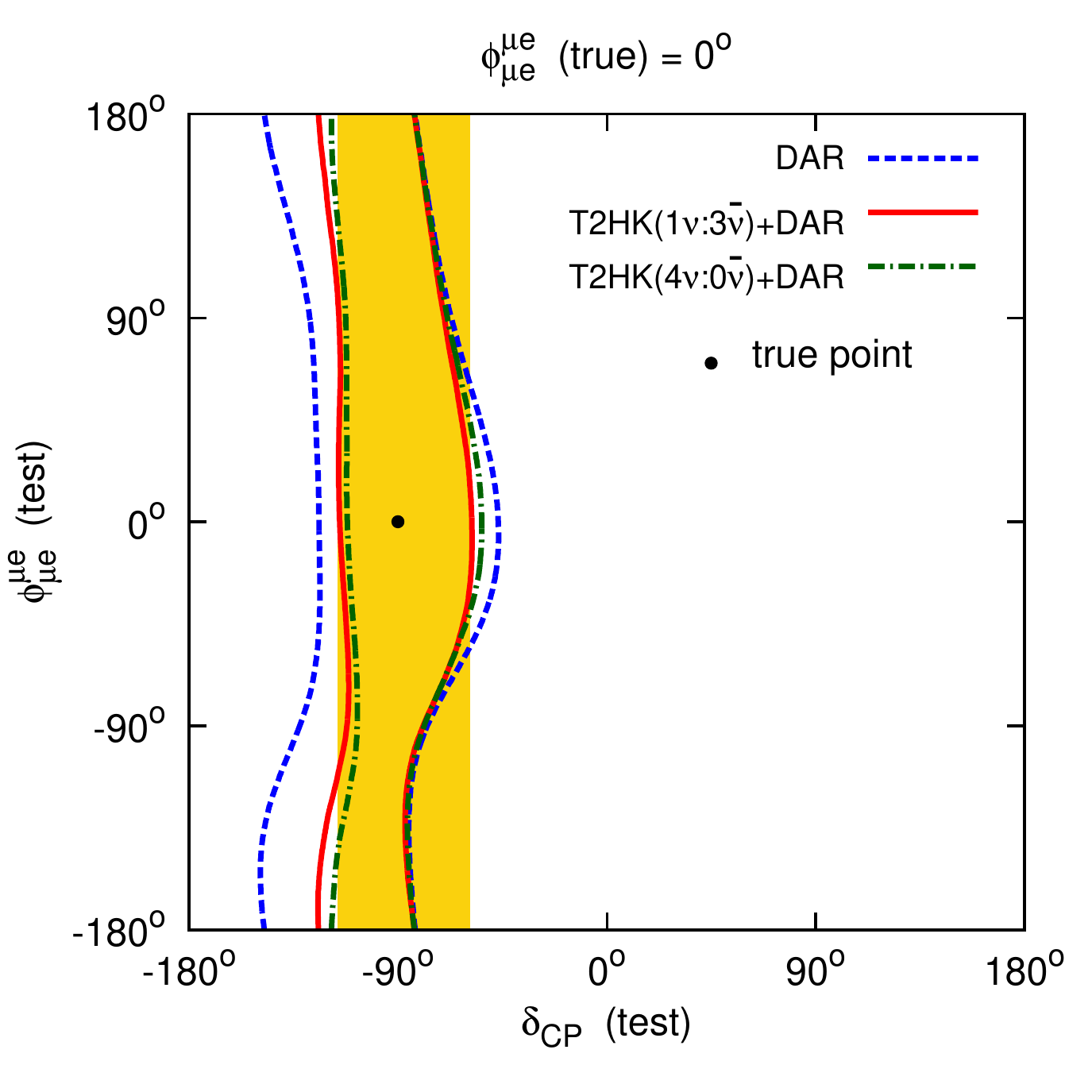}

\end{center}
\caption{\footnotesize The correlation between $\delta_{{\rm CP}}$ and  $\phi_{\mu e}$ in the test parameter plane. The true value of  $\delta_{{\rm CP}}$ is assumed to be -90$^\circ$ in all cases. However, the  true value of $\phi_{\mu e}$  is  assumed to be 90$^\circ$, 180$^\circ$, respectively in top left and right panels and -90$^\circ$, 0$^\circ$, respectively in bottom left and right panels. The neutrino mass hierarchy is assumed to be normal.}
\label{fig:correlation}
\end{figure}

Finally, we address the question of measurement of the NSI phase itself. 
We do so by finding the allowed region in the plane of $\dcp$ and $\phime$ as shown in Fig~\ref{fig:correlation} for $|\epsme|=0.01$. While the true value of $\dcp$ is always chosen to be $-90^\circ$, we choose four representative values for $\phime=-90^\circ,0^\circ,90^\circ,180^\circ$ in the four panels. The 90\%~C.L. contours are shown for \mudar\ alone, as well as in conjunction with T2HK. The synergy between \mudar\ and T2HK data helps to constrain the parameter space when $\phime=-90^\circ$ although T2HK itself is insensitive to muonic NSIs. The light band in the background represents the allowed range of $\dcp$ from T2HK+\mudar\ in the absence of NSIs. We find that for a large part of the parameter space, the precision in $\dcp$ which is given by the thickness of the allowed region does not get altered appreciably in the presence of NSIs, thereby pointing toward the robustness of the CP measurement. 
The $\phime$ precision of our setup is better for the CP-violating values of $\phime=\pm 90^\circ$ than at the CP-conserving values, $\phime= 0,180^\circ$. This is in contrast to the standard CP measurement which is better at $\dcp= 0,180^\circ$. This fact can this be explained on the basis of the top panels in Fig.~\ref{fig1}. Since the difference between probabilities in the two scenarios (SI and NSI) is highest around the maximally CP-violating values $\phime=\pm90^\circ$, the precision with which this parameter can be resolved is better.

\section{Summary and Conclusions}
\label{sec:conc}

Beyond SM scenarios that arise at a high energy scale can manifest themselves in low energy neutrino phenomena as NSIs. Thus, the search for NSIs in the neutrino sector is complementary to collider searches for new physics. At MeV scale, the NC NSIs in the neutrino propagation can be safely ignored, allowing us to focus on the CC NSIs that affect production/detection processes. 
In this work, we concentrate specifically on the muonic NSI parameters i.e., $\varepsilon^{\mu e}_{\alpha\beta}$ which uniquely impact  the production of neutrinos at a \mudar\ source. 

We compute the bounds that a \mudar\ setup can place on the magnitude of the relevant NSI parameters $\varepsilon^{\mu e}_{\mu\beta}$. For the parameter $\epsme$, the 90\%~C.L. bounds are competitive with existing bounds from tests of lepton universality for around half the range of the phase $\phime$. Addition of data from the superbeam experiment, T2HK substantially improves the bounds to better than the current ones for most of the parameter space. This is surprising, given that neutrinos produced at T2HK are not at all expected to be  affected by the muonic NSIs. This result can only be explained by invoking a synergy between T2HK and \mudar\ in constraining the standard oscillation parameters, thus lifting parameter degeneracies that hinder the measurement of NSIs. 

Next, we show that the precise measurement of $\dcp$ at the \mudar\ setup is robust even in the presence of NSIs. The precision in this parameter does not worsen considerably because of any parameter degeneracies even for the most unfavorable combinations of the two CP-violating phases -- the standard Dirac phase $\dcp$ and the nonstandard phase, $\phime$. 

Finally, we discuss correlations between  $\dcp$ and $\phime$. We find that the precision in $\phime$ is limited, and a substantial range of values of this phase can be excluded when it is close to $\pm 90^\circ$. 
In conclusion, we find that the \mudar\ setup is well suited to measure $\dcp$ not only in the standard oscillation scenario but also in the presence of CC muonic NSIs at source. In addition, this setup can also constrain the magnitude and phase of the NSI parameters depending upon its true value.

\begin{acknowledgments}
This work was initiated during the working group discussions at Workshop on High Energy Physics Phenomenology held at IIT Kanpur during December 2015 and the authors would like to thank the organizers for the vibrant atmosphere and warm hospitality during the meeting.
 S.K.R. acknowledges support by IBS (Project Code IBS-R018-D1), Basic Science Research Program through the National Research Foundation of Korea (NRF) funded by the Ministry of Education (2019R1A6A1A10073887) and NRF Strategic Research Program (NRF2017R1E1A1A01072736).  
The work of P.M. is supported by the Indian funding from University Grants Commission under the second phase of University with Potential of Excellence (UPE II) at JNU and Department of Science and Technology under DST-PURSE at JNU. P.M. would like to acknowledge partial funding from the European Union's Horizon 2020 research and innovation programme under the Marie Skodowska-Curie Grants No. 690575 and No. 674896.
\end{acknowledgments}

\bibliography{references}

%merlin.mbs apsrev4-1.bst 2010-07-25 4.21a (PWD, AO, DPC) hacked
%Control: key (0)
%Control: author (8) initials jnrlst
%Control: editor formatted (1) identically to author
%Control: production of article title (-1) disabled
%Control: page (0) single
%Control: year (1) truncated
%Control: production of eprint (0) enabled
\begin{thebibliography}{56}%
\makeatletter
\providecommand \@ifxundefined [1]{%
 \@ifx{#1\undefined}
}%
\providecommand \@ifnum [1]{%
 \ifnum #1\expandafter \@firstoftwo
 \else \expandafter \@secondoftwo
 \fi
}%
\providecommand \@ifx [1]{%
 \ifx #1\expandafter \@firstoftwo
 \else \expandafter \@secondoftwo
 \fi
}%
\providecommand \natexlab [1]{#1}%
\providecommand \enquote  [1]{``#1''}%
\providecommand \bibnamefont  [1]{#1}%
\providecommand \bibfnamefont [1]{#1}%
\providecommand \citenamefont [1]{#1}%
\providecommand \href@noop [0]{\@secondoftwo}%
\providecommand \href [0]{\begingroup \@sanitize@url \@href}%
\providecommand \@href[1]{\@@startlink{#1}\@@href}%
\providecommand \@@href[1]{\endgroup#1\@@endlink}%
\providecommand \@sanitize@url [0]{\catcode `\\12\catcode `\$12\catcode
  `\&12\catcode `\#12\catcode `\^12\catcode `\_12\catcode `\%12\relax}%
\providecommand \@@startlink[1]{}%
\providecommand \@@endlink[0]{}%
\providecommand \url  [0]{\begingroup\@sanitize@url \@url }%
\providecommand \@url [1]{\endgroup\@href {#1}{\urlprefix }}%
\providecommand \urlprefix  [0]{URL }%
\providecommand \Eprint [0]{\href }%
\providecommand \doibase [0]{http://dx.doi.org/}%
\providecommand \selectlanguage [0]{\@gobble}%
\providecommand \bibinfo  [0]{\@secondoftwo}%
\providecommand \bibfield  [0]{\@secondoftwo}%
\providecommand \translation [1]{[#1]}%
\providecommand \BibitemOpen [0]{}%
\providecommand \bibitemStop [0]{}%
\providecommand \bibitemNoStop [0]{.\EOS\space}%
\providecommand \EOS [0]{\spacefactor3000\relax}%
\providecommand \BibitemShut  [1]{\csname bibitem#1\endcsname}%
\let\auto@bib@innerbib\@empty
%</preamble>
\bibitem [{\citenamefont {Esteban}\ \emph {et~al.}(2017)\citenamefont
  {Esteban}, \citenamefont {Gonzalez-Garcia}, \citenamefont {Maltoni},
  \citenamefont {Martinez-Soler},\ and\ \citenamefont
  {Schwetz}}]{Esteban:2016qun}%
  \BibitemOpen
  \bibfield  {author} {\bibinfo {author} {\bibfnamefont {I.}~\bibnamefont
  {Esteban}}, \bibinfo {author} {\bibfnamefont {M.~C.}\ \bibnamefont
  {Gonzalez-Garcia}}, \bibinfo {author} {\bibfnamefont {M.}~\bibnamefont
  {Maltoni}}, \bibinfo {author} {\bibfnamefont {I.}~\bibnamefont
  {Martinez-Soler}}, \ and\ \bibinfo {author} {\bibfnamefont {T.}~\bibnamefont
  {Schwetz}},\ }\href {\doibase 10.1007/JHEP01(2017)087} {\bibfield  {journal}
  {\bibinfo  {journal} {JHEP}\ }\textbf {\bibinfo {volume} {01}},\ \bibinfo
  {pages} {087} (\bibinfo {year} {2017})},\ \Eprint
  {http://arxiv.org/abs/1611.01514} {arXiv:1611.01514 [hep-ph]} \BibitemShut
  {NoStop}%
%%CITATION = ARXIV:1611.01514;%%
\bibitem [{\citenamefont {Davidson}\ and\ \citenamefont
  {Sanz}(2011)}]{Davidson:2011kr}%
  \BibitemOpen
  \bibfield  {author} {\bibinfo {author} {\bibfnamefont {S.}~\bibnamefont
  {Davidson}}\ and\ \bibinfo {author} {\bibfnamefont {V.}~\bibnamefont
  {Sanz}},\ }\href {\doibase 10.1103/PhysRevD.84.113011} {\bibfield  {journal}
  {\bibinfo  {journal} {Phys. Rev.}\ }\textbf {\bibinfo {volume} {D84}},\
  \bibinfo {pages} {113011} (\bibinfo {year} {2011})},\ \Eprint
  {http://arxiv.org/abs/1108.5320} {arXiv:1108.5320 [hep-ph]} \BibitemShut
  {NoStop}%
%%CITATION = ARXIV:1108.5320;%%
\bibitem [{\citenamefont {Bhupal~Dev}\ \emph {et~al.}(2019)\citenamefont
  {Bhupal~Dev} \emph {et~al.}}]{Dev:2019anc}%
  \BibitemOpen
  \bibfield  {author} {\bibinfo {author} {\bibfnamefont {P.~S.}\ \bibnamefont
  {Bhupal~Dev}} \emph {et~al.},\ }in\ \href
  {http://lss.fnal.gov/archive/2019/conf/fermilab-conf-19-299-t.pdf} {\emph
  {\bibinfo {booktitle} {{NTN Workshop on Neutrino Non-Standard Interactions St
  Louis, MO, USA, May 29-31, 2019}}}}\ (\bibinfo {year} {2019})\ \Eprint
  {http://arxiv.org/abs/1907.00991} {arXiv:1907.00991 [hep-ph]} \BibitemShut
  {NoStop}%
%%CITATION = ARXIV:1907.00991;%%
\bibitem [{\citenamefont {Wolfenstein}(1978)}]{Wolfenstein:1977ue}%
  \BibitemOpen
  \bibfield  {author} {\bibinfo {author} {\bibfnamefont {L.}~\bibnamefont
  {Wolfenstein}},\ }\href {\doibase 10.1103/PhysRevD.17.2369} {\bibfield
  {journal} {\bibinfo  {journal} {Phys. Rev.}\ }\textbf {\bibinfo {volume}
  {D17}},\ \bibinfo {pages} {2369} (\bibinfo {year} {1978})}\BibitemShut
  {NoStop}%
%%CITATION = PHRVA,D17,2369;%%
\bibitem [{\citenamefont {Miranda}\ and\ \citenamefont
  {Nunokawa}(2015)}]{Miranda:2015dra}%
  \BibitemOpen
  \bibfield  {author} {\bibinfo {author} {\bibfnamefont {O.~G.}\ \bibnamefont
  {Miranda}}\ and\ \bibinfo {author} {\bibfnamefont {H.}~\bibnamefont
  {Nunokawa}},\ }\href {\doibase 10.1088/1367-2630/17/9/095002} {\bibfield
  {journal} {\bibinfo  {journal} {New J. Phys.}\ }\textbf {\bibinfo {volume}
  {17}},\ \bibinfo {pages} {095002} (\bibinfo {year} {2015})},\ \Eprint
  {http://arxiv.org/abs/1505.06254} {arXiv:1505.06254 [hep-ph]} \BibitemShut
  {NoStop}%
%%CITATION = ARXIV:1505.06254;%%
\bibitem [{\citenamefont {Ohlsson}(2013)}]{Ohlsson:2012kf}%
  \BibitemOpen
  \bibfield  {author} {\bibinfo {author} {\bibfnamefont {T.}~\bibnamefont
  {Ohlsson}},\ }\href {\doibase 10.1088/0034-4885/76/4/044201} {\bibfield
  {journal} {\bibinfo  {journal} {Rept. Prog. Phys.}\ }\textbf {\bibinfo
  {volume} {76}},\ \bibinfo {pages} {044201} (\bibinfo {year} {2013})},\
  \Eprint {http://arxiv.org/abs/1209.2710} {arXiv:1209.2710 [hep-ph]}
  \BibitemShut {NoStop}%
%%CITATION = ARXIV:1209.2710;%%
\bibitem [{\citenamefont {Farzan}\ and\ \citenamefont
  {Tortola}(2018)}]{Farzan:2017xzy}%
  \BibitemOpen
  \bibfield  {author} {\bibinfo {author} {\bibfnamefont {Y.}~\bibnamefont
  {Farzan}}\ and\ \bibinfo {author} {\bibfnamefont {M.}~\bibnamefont
  {Tortola}},\ }\href {\doibase 10.3389/fphy.2018.00010} {\bibfield  {journal}
  {\bibinfo  {journal} {Front.in Phys.}\ }\textbf {\bibinfo {volume} {6}},\
  \bibinfo {pages} {10} (\bibinfo {year} {2018})},\ \Eprint
  {http://arxiv.org/abs/1710.09360} {arXiv:1710.09360 [hep-ph]} \BibitemShut
  {NoStop}%
%%CITATION = ARXIV:1710.09360;%%
\bibitem [{\citenamefont {Pilaftsis}\ and\ \citenamefont
  {Underwood}(2005)}]{Pilaftsis:2005rv}%
  \BibitemOpen
  \bibfield  {author} {\bibinfo {author} {\bibfnamefont {A.}~\bibnamefont
  {Pilaftsis}}\ and\ \bibinfo {author} {\bibfnamefont {T.~E.~J.}\ \bibnamefont
  {Underwood}},\ }\href {\doibase 10.1103/PhysRevD.72.113001} {\bibfield
  {journal} {\bibinfo  {journal} {Phys. Rev.}\ }\textbf {\bibinfo {volume}
  {D72}},\ \bibinfo {pages} {113001} (\bibinfo {year} {2005})},\ \Eprint
  {http://arxiv.org/abs/hep-ph/0506107} {arXiv:hep-ph/0506107 [hep-ph]}
  \BibitemShut {NoStop}%
%%CITATION = HEP-PH/0506107;%%
\bibitem [{\citenamefont {Barbieri}\ \emph {et~al.}(1990)\citenamefont
  {Barbieri}, \citenamefont {Guzzo}, \citenamefont {Masiero},\ and\
  \citenamefont {Tommasini}}]{BARBIERI1990251}%
  \BibitemOpen
  \bibfield  {author} {\bibinfo {author} {\bibfnamefont {R.}~\bibnamefont
  {Barbieri}}, \bibinfo {author} {\bibfnamefont {M.}~\bibnamefont {Guzzo}},
  \bibinfo {author} {\bibfnamefont {A.}~\bibnamefont {Masiero}}, \ and\
  \bibinfo {author} {\bibfnamefont {D.}~\bibnamefont {Tommasini}},\ }\href
  {\doibase https://doi.org/10.1016/0370-2693(90)90869-8} {\bibfield  {journal}
  {\bibinfo  {journal} {Physics Letters B}\ }\textbf {\bibinfo {volume}
  {252}},\ \bibinfo {pages} {251 } (\bibinfo {year} {1990})}\BibitemShut
  {NoStop}%
\bibitem [{\citenamefont {Babu}\ and\ \citenamefont
  {Mohapatra}(1989)}]{PhysRevLett.63.228}%
  \BibitemOpen
  \bibfield  {author} {\bibinfo {author} {\bibfnamefont {K.~S.}\ \bibnamefont
  {Babu}}\ and\ \bibinfo {author} {\bibfnamefont {R.~N.}\ \bibnamefont
  {Mohapatra}},\ }\href {\doibase 10.1103/PhysRevLett.63.228} {\bibfield
  {journal} {\bibinfo  {journal} {Phys. Rev. Lett.}\ }\textbf {\bibinfo
  {volume} {63}},\ \bibinfo {pages} {228} (\bibinfo {year} {1989})}\BibitemShut
  {NoStop}%
\bibitem [{\citenamefont {Choudhury}\ and\ \citenamefont
  {Sarkar}(1990)}]{CHOUDHURY1990113}%
  \BibitemOpen
  \bibfield  {author} {\bibinfo {author} {\bibfnamefont {D.}~\bibnamefont
  {Choudhury}}\ and\ \bibinfo {author} {\bibfnamefont {U.}~\bibnamefont
  {Sarkar}},\ }\href {\doibase https://doi.org/10.1016/0370-2693(90)90105-F}
  {\bibfield  {journal} {\bibinfo  {journal} {Physics Letters B}\ }\textbf
  {\bibinfo {volume} {235}},\ \bibinfo {pages} {113 } (\bibinfo {year}
  {1990})}\BibitemShut {NoStop}%
\bibitem [{\citenamefont {Gonzalez-Garcia}\ \emph {et~al.}(2018)\citenamefont
  {Gonzalez-Garcia}, \citenamefont {Maltoni}, \citenamefont {Perez-Gonzalez},\
  and\ \citenamefont {Zukanovich~Funchal}}]{Gonzalez-Garcia:2018dep}%
  \BibitemOpen
  \bibfield  {author} {\bibinfo {author} {\bibfnamefont {M.~C.}\ \bibnamefont
  {Gonzalez-Garcia}}, \bibinfo {author} {\bibfnamefont {M.}~\bibnamefont
  {Maltoni}}, \bibinfo {author} {\bibfnamefont {Y.~F.}\ \bibnamefont
  {Perez-Gonzalez}}, \ and\ \bibinfo {author} {\bibfnamefont {R.}~\bibnamefont
  {Zukanovich~Funchal}},\ }\href {\doibase 10.1007/JHEP07(2018)019} {\bibfield
  {journal} {\bibinfo  {journal} {JHEP}\ }\textbf {\bibinfo {volume} {07}},\
  \bibinfo {pages} {019} (\bibinfo {year} {2018})},\ \Eprint
  {http://arxiv.org/abs/1803.03650} {arXiv:1803.03650 [hep-ph]} \BibitemShut
  {NoStop}%
%%CITATION = ARXIV:1803.03650;%%
\bibitem [{\citenamefont {Davidson}\ \emph {et~al.}(2003)\citenamefont
  {Davidson}, \citenamefont {Pena-Garay}, \citenamefont {Rius},\ and\
  \citenamefont {Santamaria}}]{Davidson:2003ha}%
  \BibitemOpen
  \bibfield  {author} {\bibinfo {author} {\bibfnamefont {S.}~\bibnamefont
  {Davidson}}, \bibinfo {author} {\bibfnamefont {C.}~\bibnamefont
  {Pena-Garay}}, \bibinfo {author} {\bibfnamefont {N.}~\bibnamefont {Rius}}, \
  and\ \bibinfo {author} {\bibfnamefont {A.}~\bibnamefont {Santamaria}},\
  }\href {\doibase 10.1088/1126-6708/2003/03/011} {\bibfield  {journal}
  {\bibinfo  {journal} {JHEP}\ }\textbf {\bibinfo {volume} {03}},\ \bibinfo
  {pages} {011} (\bibinfo {year} {2003})},\ \Eprint
  {http://arxiv.org/abs/hep-ph/0302093} {arXiv:hep-ph/0302093 [hep-ph]}
  \BibitemShut {NoStop}%
%%CITATION = HEP-PH/0302093;%%
\bibitem [{\citenamefont {Biggio}\ \emph {et~al.}(2009)\citenamefont {Biggio},
  \citenamefont {Blennow},\ and\ \citenamefont
  {Fernandez-Martinez}}]{Biggio:2009nt}%
  \BibitemOpen
  \bibfield  {author} {\bibinfo {author} {\bibfnamefont {C.}~\bibnamefont
  {Biggio}}, \bibinfo {author} {\bibfnamefont {M.}~\bibnamefont {Blennow}}, \
  and\ \bibinfo {author} {\bibfnamefont {E.}~\bibnamefont
  {Fernandez-Martinez}},\ }\href {\doibase 10.1088/1126-6708/2009/08/090}
  {\bibfield  {journal} {\bibinfo  {journal} {JHEP}\ }\textbf {\bibinfo
  {volume} {0908}},\ \bibinfo {pages} {090} (\bibinfo {year} {2009})},\ \Eprint
  {http://arxiv.org/abs/0907.0097} {arXiv:0907.0097 [hep-ph]} \BibitemShut
  {NoStop}%
%%CITATION = ARXIV:0907.0097;%%
\bibitem [{\citenamefont {Masud}\ \emph {et~al.}(2016)\citenamefont {Masud},
  \citenamefont {Chatterjee},\ and\ \citenamefont {Mehta}}]{Masud:2015xva}%
  \BibitemOpen
  \bibfield  {author} {\bibinfo {author} {\bibfnamefont {M.}~\bibnamefont
  {Masud}}, \bibinfo {author} {\bibfnamefont {A.}~\bibnamefont {Chatterjee}}, \
  and\ \bibinfo {author} {\bibfnamefont {P.}~\bibnamefont {Mehta}},\ }\href
  {\doibase 10.1088/0954-3899/43/9/095005/meta, 10.1088/0954-3899/43/9/095005}
  {\bibfield  {journal} {\bibinfo  {journal} {J. Phys.}\ }\textbf {\bibinfo
  {volume} {G43}},\ \bibinfo {pages} {095005} (\bibinfo {year} {2016})},\
  \Eprint {http://arxiv.org/abs/1510.08261} {arXiv:1510.08261 [hep-ph]}
  \BibitemShut {NoStop}%
%%CITATION = ARXIV:1510.08261;%%
\bibitem [{\citenamefont {Coloma}(2016)}]{Coloma:2015kiu}%
  \BibitemOpen
  \bibfield  {author} {\bibinfo {author} {\bibfnamefont {P.}~\bibnamefont
  {Coloma}},\ }\href {\doibase 10.1007/JHEP03(2016)016} {\bibfield  {journal}
  {\bibinfo  {journal} {JHEP}\ }\textbf {\bibinfo {volume} {03}},\ \bibinfo
  {pages} {016} (\bibinfo {year} {2016})},\ \Eprint
  {http://arxiv.org/abs/1511.06357} {arXiv:1511.06357 [hep-ph]} \BibitemShut
  {NoStop}%
%%CITATION = ARXIV:1511.06357;%%
\bibitem [{\citenamefont {de~Gouvea}\ and\ \citenamefont
  {Kelly}(2015)}]{deGouvea:2015ndi}%
  \BibitemOpen
  \bibfield  {author} {\bibinfo {author} {\bibfnamefont {A.}~\bibnamefont
  {de~Gouvea}}\ and\ \bibinfo {author} {\bibfnamefont {K.~J.}\ \bibnamefont
  {Kelly}},\ }\href@noop {} {\  (\bibinfo {year} {2015})},\ \Eprint
  {http://arxiv.org/abs/1511.05562} {arXiv:1511.05562 [hep-ph]} \BibitemShut
  {NoStop}%
%%CITATION = ARXIV:1511.05562;%%
\bibitem [{\citenamefont {Blennow}\ \emph {et~al.}(2016)\citenamefont
  {Blennow}, \citenamefont {Choubey}, \citenamefont {Ohlsson}, \citenamefont
  {Pramanik},\ and\ \citenamefont {Raut}}]{Blennow:2016etl}%
  \BibitemOpen
  \bibfield  {author} {\bibinfo {author} {\bibfnamefont {M.}~\bibnamefont
  {Blennow}}, \bibinfo {author} {\bibfnamefont {S.}~\bibnamefont {Choubey}},
  \bibinfo {author} {\bibfnamefont {T.}~\bibnamefont {Ohlsson}}, \bibinfo
  {author} {\bibfnamefont {D.}~\bibnamefont {Pramanik}}, \ and\ \bibinfo
  {author} {\bibfnamefont {S.~K.}\ \bibnamefont {Raut}},\ }\href {\doibase
  10.1007/JHEP08(2016)090} {\bibfield  {journal} {\bibinfo  {journal} {JHEP}\
  }\textbf {\bibinfo {volume} {08}},\ \bibinfo {pages} {090} (\bibinfo {year}
  {2016})},\ \Eprint {http://arxiv.org/abs/1606.08851} {arXiv:1606.08851
  [hep-ph]} \BibitemShut {NoStop}%
%%CITATION = ARXIV:1606.08851;%%
\bibitem [{\citenamefont {Deepthi}\ \emph {et~al.}(2018)\citenamefont
  {Deepthi}, \citenamefont {Goswami},\ and\ \citenamefont
  {Nath}}]{Deepthi:2017gxg}%
  \BibitemOpen
  \bibfield  {author} {\bibinfo {author} {\bibfnamefont {K.~N.}\ \bibnamefont
  {Deepthi}}, \bibinfo {author} {\bibfnamefont {S.}~\bibnamefont {Goswami}}, \
  and\ \bibinfo {author} {\bibfnamefont {N.}~\bibnamefont {Nath}},\ }\href
  {\doibase 10.1016/j.nuclphysb.2018.09.004} {\bibfield  {journal} {\bibinfo
  {journal} {Nucl. Phys.}\ }\textbf {\bibinfo {volume} {B936}},\ \bibinfo
  {pages} {91} (\bibinfo {year} {2018})},\ \Eprint
  {http://arxiv.org/abs/1711.04840} {arXiv:1711.04840 [hep-ph]} \BibitemShut
  {NoStop}%
%%CITATION = ARXIV:1711.04840;%%
\bibitem [{\citenamefont {Masud}\ \emph {et~al.}(2019)\citenamefont {Masud},
  \citenamefont {Roy},\ and\ \citenamefont {Mehta}}]{Masud:2018pig}%
  \BibitemOpen
  \bibfield  {author} {\bibinfo {author} {\bibfnamefont {M.}~\bibnamefont
  {Masud}}, \bibinfo {author} {\bibfnamefont {S.}~\bibnamefont {Roy}}, \ and\
  \bibinfo {author} {\bibfnamefont {P.}~\bibnamefont {Mehta}},\ }\href
  {\doibase 10.1103/PhysRevD.99.115032} {\bibfield  {journal} {\bibinfo
  {journal} {Phys. Rev.}\ }\textbf {\bibinfo {volume} {D99}},\ \bibinfo {pages}
  {115032} (\bibinfo {year} {2019})},\ \Eprint
  {http://arxiv.org/abs/1812.10290} {arXiv:1812.10290 [hep-ph]} \BibitemShut
  {NoStop}%
%%CITATION = ARXIV:1812.10290;%%
\bibitem [{\citenamefont {Kopp}\ \emph {et~al.}(2008)\citenamefont {Kopp},
  \citenamefont {Lindner}, \citenamefont {Ota},\ and\ \citenamefont
  {Sato}}]{Kopp:2007ne}%
  \BibitemOpen
  \bibfield  {author} {\bibinfo {author} {\bibfnamefont {J.}~\bibnamefont
  {Kopp}}, \bibinfo {author} {\bibfnamefont {M.}~\bibnamefont {Lindner}},
  \bibinfo {author} {\bibfnamefont {T.}~\bibnamefont {Ota}}, \ and\ \bibinfo
  {author} {\bibfnamefont {J.}~\bibnamefont {Sato}},\ }\href {\doibase
  10.1103/PhysRevD.77.013007} {\bibfield  {journal} {\bibinfo  {journal} {Phys.
  Rev.}\ }\textbf {\bibinfo {volume} {D77}},\ \bibinfo {pages} {013007}
  (\bibinfo {year} {2008})},\ \Eprint {http://arxiv.org/abs/0708.0152}
  {arXiv:0708.0152 [hep-ph]} \BibitemShut {NoStop}%
%%CITATION = ARXIV:0708.0152;%%
\bibitem [{\citenamefont {Das}\ and\ \citenamefont
  {Pulido}(2011)}]{Das:2010sd}%
  \BibitemOpen
  \bibfield  {author} {\bibinfo {author} {\bibfnamefont {C.~R.}\ \bibnamefont
  {Das}}\ and\ \bibinfo {author} {\bibfnamefont {J.}~\bibnamefont {Pulido}},\
  }\href {\doibase 10.1103/PhysRevD.83.053009} {\bibfield  {journal} {\bibinfo
  {journal} {Phys. Rev.}\ }\textbf {\bibinfo {volume} {D83}},\ \bibinfo {pages}
  {053009} (\bibinfo {year} {2011})},\ \Eprint {http://arxiv.org/abs/1007.2167}
  {arXiv:1007.2167 [hep-ph]} \BibitemShut {NoStop}%
%%CITATION = ARXIV:1007.2167;%%
\bibitem [{\citenamefont {Chatterjee}\ \emph {et~al.}(2016)\citenamefont
  {Chatterjee}, \citenamefont {Mehta}, \citenamefont {Choudhury},\ and\
  \citenamefont {Gandhi}}]{Chatterjee:2014gxa}%
  \BibitemOpen
  \bibfield  {author} {\bibinfo {author} {\bibfnamefont {A.}~\bibnamefont
  {Chatterjee}}, \bibinfo {author} {\bibfnamefont {P.}~\bibnamefont {Mehta}},
  \bibinfo {author} {\bibfnamefont {D.}~\bibnamefont {Choudhury}}, \ and\
  \bibinfo {author} {\bibfnamefont {R.}~\bibnamefont {Gandhi}},\ }\href
  {\doibase 10.1103/PhysRevD.93.093017} {\bibfield  {journal} {\bibinfo
  {journal} {Phys. Rev.}\ }\textbf {\bibinfo {volume} {D93}},\ \bibinfo {pages}
  {093017} (\bibinfo {year} {2016})},\ \Eprint {http://arxiv.org/abs/1409.8472}
  {arXiv:1409.8472 [hep-ph]} \BibitemShut {NoStop}%
%%CITATION = ARXIV:1409.8472;%%
\bibitem [{\citenamefont {Choubey}\ \emph {et~al.}(2015)\citenamefont
  {Choubey}, \citenamefont {Ghosh}, \citenamefont {Ohlsson},\ and\
  \citenamefont {Tiwari}}]{Choubey:2015xha}%
  \BibitemOpen
  \bibfield  {author} {\bibinfo {author} {\bibfnamefont {S.}~\bibnamefont
  {Choubey}}, \bibinfo {author} {\bibfnamefont {A.}~\bibnamefont {Ghosh}},
  \bibinfo {author} {\bibfnamefont {T.}~\bibnamefont {Ohlsson}}, \ and\
  \bibinfo {author} {\bibfnamefont {D.}~\bibnamefont {Tiwari}},\ }\href
  {\doibase 10.1007/JHEP12(2015)126} {\bibfield  {journal} {\bibinfo  {journal}
  {JHEP}\ }\textbf {\bibinfo {volume} {12}},\ \bibinfo {pages} {126} (\bibinfo
  {year} {2015})},\ \Eprint {http://arxiv.org/abs/1507.02211} {arXiv:1507.02211
  [hep-ph]} \BibitemShut {NoStop}%
%%CITATION = ARXIV:1507.02211;%%
\bibitem [{\citenamefont {Blennow}\ \emph {et~al.}(2015)\citenamefont
  {Blennow}, \citenamefont {Choubey}, \citenamefont {Ohlsson},\ and\
  \citenamefont {Raut}}]{Blennow:2015nxa}%
  \BibitemOpen
  \bibfield  {author} {\bibinfo {author} {\bibfnamefont {M.}~\bibnamefont
  {Blennow}}, \bibinfo {author} {\bibfnamefont {S.}~\bibnamefont {Choubey}},
  \bibinfo {author} {\bibfnamefont {T.}~\bibnamefont {Ohlsson}}, \ and\
  \bibinfo {author} {\bibfnamefont {S.~K.}\ \bibnamefont {Raut}},\ }\href
  {\doibase 10.1007/JHEP09(2015)096} {\bibfield  {journal} {\bibinfo  {journal}
  {JHEP}\ }\textbf {\bibinfo {volume} {09}},\ \bibinfo {pages} {096} (\bibinfo
  {year} {2015})},\ \Eprint {http://arxiv.org/abs/1507.02868} {arXiv:1507.02868
  [hep-ph]} \BibitemShut {NoStop}%
%%CITATION = ARXIV:1507.02868;%%
\bibitem [{\citenamefont {Bakhti}\ \emph {et~al.}(2017)\citenamefont {Bakhti},
  \citenamefont {Khan},\ and\ \citenamefont {Wang}}]{Bakhti:2016gic}%
  \BibitemOpen
  \bibfield  {author} {\bibinfo {author} {\bibfnamefont {P.}~\bibnamefont
  {Bakhti}}, \bibinfo {author} {\bibfnamefont {A.~N.}\ \bibnamefont {Khan}}, \
  and\ \bibinfo {author} {\bibfnamefont {W.}~\bibnamefont {Wang}},\ }\href
  {\doibase 10.1088/1361-6471/aa9098} {\bibfield  {journal} {\bibinfo
  {journal} {J. Phys.}\ }\textbf {\bibinfo {volume} {G44}},\ \bibinfo {pages}
  {125001} (\bibinfo {year} {2017})},\ \Eprint
  {http://arxiv.org/abs/1607.00065} {arXiv:1607.00065 [hep-ph]} \BibitemShut
  {NoStop}%
%%CITATION = ARXIV:1607.00065;%%
\bibitem [{\citenamefont {Blennow}\ and\ \citenamefont
  {Meloni}(2009)}]{Blennow:2009rp}%
  \BibitemOpen
  \bibfield  {author} {\bibinfo {author} {\bibfnamefont {M.}~\bibnamefont
  {Blennow}}\ and\ \bibinfo {author} {\bibfnamefont {D.}~\bibnamefont
  {Meloni}},\ }\href {\doibase 10.1103/PhysRevD.80.065009} {\bibfield
  {journal} {\bibinfo  {journal} {Phys. Rev.}\ }\textbf {\bibinfo {volume}
  {D80}},\ \bibinfo {pages} {065009} (\bibinfo {year} {2009})},\ \Eprint
  {http://arxiv.org/abs/0901.2110} {arXiv:0901.2110 [hep-ph]} \BibitemShut
  {NoStop}%
%%CITATION = ARXIV:0901.2110;%%
\bibitem [{\citenamefont {Meloni}\ \emph {et~al.}(2010)\citenamefont {Meloni},
  \citenamefont {Ohlsson}, \citenamefont {Winter},\ and\ \citenamefont
  {Zhang}}]{Meloni:2009cg}%
  \BibitemOpen
  \bibfield  {author} {\bibinfo {author} {\bibfnamefont {D.}~\bibnamefont
  {Meloni}}, \bibinfo {author} {\bibfnamefont {T.}~\bibnamefont {Ohlsson}},
  \bibinfo {author} {\bibfnamefont {W.}~\bibnamefont {Winter}}, \ and\ \bibinfo
  {author} {\bibfnamefont {H.}~\bibnamefont {Zhang}},\ }\href {\doibase
  10.1007/JHEP04(2010)041} {\bibfield  {journal} {\bibinfo  {journal} {JHEP}\
  }\textbf {\bibinfo {volume} {04}},\ \bibinfo {pages} {041} (\bibinfo {year}
  {2010})},\ \Eprint {http://arxiv.org/abs/0912.2735} {arXiv:0912.2735
  [hep-ph]} \BibitemShut {NoStop}%
%%CITATION = ARXIV:0912.2735;%%
\bibitem [{\citenamefont {Akhmedov}\ and\ \citenamefont
  {Schwetz}(2010)}]{Akhmedov:2010vy}%
  \BibitemOpen
  \bibfield  {author} {\bibinfo {author} {\bibfnamefont {E.}~\bibnamefont
  {Akhmedov}}\ and\ \bibinfo {author} {\bibfnamefont {T.}~\bibnamefont
  {Schwetz}},\ }\href {\doibase 10.1007/JHEP10(2010)115} {\bibfield  {journal}
  {\bibinfo  {journal} {JHEP}\ }\textbf {\bibinfo {volume} {10}},\ \bibinfo
  {pages} {115} (\bibinfo {year} {2010})},\ \Eprint
  {http://arxiv.org/abs/1007.4171} {arXiv:1007.4171 [hep-ph]} \BibitemShut
  {NoStop}%
%%CITATION = ARXIV:1007.4171;%%
\bibitem [{\citenamefont {Bolanos}\ \emph {et~al.}(2009)\citenamefont
  {Bolanos}, \citenamefont {Miranda}, \citenamefont {Palazzo}, \citenamefont
  {Tortola},\ and\ \citenamefont {Valle}}]{Bolanos:2008km}%
  \BibitemOpen
  \bibfield  {author} {\bibinfo {author} {\bibfnamefont {A.}~\bibnamefont
  {Bolanos}}, \bibinfo {author} {\bibfnamefont {O.~G.}\ \bibnamefont
  {Miranda}}, \bibinfo {author} {\bibfnamefont {A.}~\bibnamefont {Palazzo}},
  \bibinfo {author} {\bibfnamefont {M.~A.}\ \bibnamefont {Tortola}}, \ and\
  \bibinfo {author} {\bibfnamefont {J.~W.~F.}\ \bibnamefont {Valle}},\ }\href
  {\doibase 10.1103/PhysRevD.79.113012} {\bibfield  {journal} {\bibinfo
  {journal} {Phys. Rev.}\ }\textbf {\bibinfo {volume} {D79}},\ \bibinfo {pages}
  {113012} (\bibinfo {year} {2009})},\ \Eprint {http://arxiv.org/abs/0812.4417}
  {arXiv:0812.4417 [hep-ph]} \BibitemShut {NoStop}%
%%CITATION = ARXIV:0812.4417;%%
\bibitem [{\citenamefont {Leitner}\ \emph {et~al.}(2011)\citenamefont
  {Leitner}, \citenamefont {Malinsky}, \citenamefont {Roskovec},\ and\
  \citenamefont {Zhang}}]{Leitner:2011aa}%
  \BibitemOpen
  \bibfield  {author} {\bibinfo {author} {\bibfnamefont {R.}~\bibnamefont
  {Leitner}}, \bibinfo {author} {\bibfnamefont {M.}~\bibnamefont {Malinsky}},
  \bibinfo {author} {\bibfnamefont {B.}~\bibnamefont {Roskovec}}, \ and\
  \bibinfo {author} {\bibfnamefont {H.}~\bibnamefont {Zhang}},\ }\href
  {\doibase 10.1007/JHEP12(2011)001} {\bibfield  {journal} {\bibinfo  {journal}
  {JHEP}\ }\textbf {\bibinfo {volume} {12}},\ \bibinfo {pages} {001} (\bibinfo
  {year} {2011})},\ \Eprint {http://arxiv.org/abs/1105.5580} {arXiv:1105.5580
  [hep-ph]} \BibitemShut {NoStop}%
%%CITATION = ARXIV:1105.5580;%%
\bibitem [{\citenamefont {Khan}\ \emph {et~al.}(2013)\citenamefont {Khan},
  \citenamefont {McKay},\ and\ \citenamefont {Tahir}}]{Khan:2013hva}%
  \BibitemOpen
  \bibfield  {author} {\bibinfo {author} {\bibfnamefont {A.~N.}\ \bibnamefont
  {Khan}}, \bibinfo {author} {\bibfnamefont {D.~W.}\ \bibnamefont {McKay}}, \
  and\ \bibinfo {author} {\bibfnamefont {F.}~\bibnamefont {Tahir}},\ }\href
  {\doibase 10.1103/PhysRevD.88.113006} {\bibfield  {journal} {\bibinfo
  {journal} {Phys. Rev.}\ }\textbf {\bibinfo {volume} {D88}},\ \bibinfo {pages}
  {113006} (\bibinfo {year} {2013})},\ \Eprint {http://arxiv.org/abs/1305.4350}
  {arXiv:1305.4350 [hep-ph]} \BibitemShut {NoStop}%
%%CITATION = ARXIV:1305.4350;%%
\bibitem [{\citenamefont {Girardi}\ \emph {et~al.}(2014)\citenamefont
  {Girardi}, \citenamefont {Meloni},\ and\ \citenamefont
  {Petcov}}]{Girardi:2014kca}%
  \BibitemOpen
  \bibfield  {author} {\bibinfo {author} {\bibfnamefont {I.}~\bibnamefont
  {Girardi}}, \bibinfo {author} {\bibfnamefont {D.}~\bibnamefont {Meloni}}, \
  and\ \bibinfo {author} {\bibfnamefont {S.~T.}\ \bibnamefont {Petcov}},\
  }\href {\doibase 10.1016/j.nuclphysb.2014.06.014} {\bibfield  {journal}
  {\bibinfo  {journal} {Nucl. Phys.}\ }\textbf {\bibinfo {volume} {B886}},\
  \bibinfo {pages} {31} (\bibinfo {year} {2014})},\ \Eprint
  {http://arxiv.org/abs/1405.0416} {arXiv:1405.0416 [hep-ph]} \BibitemShut
  {NoStop}%
%%CITATION = ARXIV:1405.0416;%%
\bibitem [{\citenamefont {Khan}\ \emph {et~al.}(2014)\citenamefont {Khan},
  \citenamefont {McKay},\ and\ \citenamefont {Tahir}}]{Khan:2014zwa}%
  \BibitemOpen
  \bibfield  {author} {\bibinfo {author} {\bibfnamefont {A.~N.}\ \bibnamefont
  {Khan}}, \bibinfo {author} {\bibfnamefont {D.~W.}\ \bibnamefont {McKay}}, \
  and\ \bibinfo {author} {\bibfnamefont {F.}~\bibnamefont {Tahir}},\ }\href
  {\doibase 10.1103/PhysRevD.90.053008} {\bibfield  {journal} {\bibinfo
  {journal} {Phys. Rev.}\ }\textbf {\bibinfo {volume} {D90}},\ \bibinfo {pages}
  {053008} (\bibinfo {year} {2014})},\ \Eprint {http://arxiv.org/abs/1407.4263}
  {arXiv:1407.4263 [hep-ph]} \BibitemShut {NoStop}%
%%CITATION = ARXIV:1407.4263;%%
\bibitem [{\citenamefont {Agarwalla}\ \emph {et~al.}(2015)\citenamefont
  {Agarwalla}, \citenamefont {Bagchi}, \citenamefont {Forero},\ and\
  \citenamefont {Tórtola}}]{Agarwalla:2014bsa}%
  \BibitemOpen
  \bibfield  {author} {\bibinfo {author} {\bibfnamefont {S.~K.}\ \bibnamefont
  {Agarwalla}}, \bibinfo {author} {\bibfnamefont {P.}~\bibnamefont {Bagchi}},
  \bibinfo {author} {\bibfnamefont {D.~V.}\ \bibnamefont {Forero}}, \ and\
  \bibinfo {author} {\bibfnamefont {M.}~\bibnamefont {Tórtola}},\ }\href
  {\doibase 10.1007/JHEP07(2015)060} {\bibfield  {journal} {\bibinfo  {journal}
  {JHEP}\ }\textbf {\bibinfo {volume} {07}},\ \bibinfo {pages} {060} (\bibinfo
  {year} {2015})},\ \Eprint {http://arxiv.org/abs/1412.1064} {arXiv:1412.1064
  [hep-ph]} \BibitemShut {NoStop}%
%%CITATION = ARXIV:1412.1064;%%
\bibitem [{\citenamefont {Khan}\ and\ \citenamefont
  {McKay}(2017)}]{Khan:2017oxw}%
  \BibitemOpen
  \bibfield  {author} {\bibinfo {author} {\bibfnamefont {A.~N.}\ \bibnamefont
  {Khan}}\ and\ \bibinfo {author} {\bibfnamefont {D.~W.}\ \bibnamefont
  {McKay}},\ }\href {\doibase 10.1007/JHEP07(2017)143} {\bibfield  {journal}
  {\bibinfo  {journal} {JHEP}\ }\textbf {\bibinfo {volume} {07}},\ \bibinfo
  {pages} {143} (\bibinfo {year} {2017})},\ \Eprint
  {http://arxiv.org/abs/1704.06222} {arXiv:1704.06222 [hep-ph]} \BibitemShut
  {NoStop}%
%%CITATION = ARXIV:1704.06222;%%
\bibitem [{\citenamefont {Bennett}\ \emph {et~al.}(2006)\citenamefont {Bennett}
  \emph {et~al.}}]{Bennett:2006fi}%
  \BibitemOpen
  \bibfield  {author} {\bibinfo {author} {\bibfnamefont {G.~W.}\ \bibnamefont
  {Bennett}} \emph {et~al.} (\bibinfo {collaboration} {Muon g-2}),\ }\href
  {\doibase 10.1103/PhysRevD.73.072003} {\bibfield  {journal} {\bibinfo
  {journal} {Phys. Rev.}\ }\textbf {\bibinfo {volume} {D73}},\ \bibinfo {pages}
  {072003} (\bibinfo {year} {2006})},\ \Eprint
  {http://arxiv.org/abs/hep-ex/0602035} {arXiv:hep-ex/0602035 [hep-ex]}
  \BibitemShut {NoStop}%
%%CITATION = HEP-EX/0602035;%%
\bibitem [{\citenamefont {Albrecht}\ \emph {et~al.}(2018)\citenamefont
  {Albrecht}, \citenamefont {Reichert},\ and\ \citenamefont {van
  Dyk}}]{Albrecht:2018vsa}%
  \BibitemOpen
  \bibfield  {author} {\bibinfo {author} {\bibfnamefont {J.}~\bibnamefont
  {Albrecht}}, \bibinfo {author} {\bibfnamefont {S.}~\bibnamefont {Reichert}},
  \ and\ \bibinfo {author} {\bibfnamefont {D.}~\bibnamefont {van Dyk}},\ }\href
  {\doibase 10.1142/S0217751X18300168} {\bibfield  {journal} {\bibinfo
  {journal} {Int. J. Mod. Phys.}\ }\textbf {\bibinfo {volume} {A33}},\ \bibinfo
  {pages} {1830016} (\bibinfo {year} {2018})},\ \Eprint
  {http://arxiv.org/abs/1806.05010} {arXiv:1806.05010 [hep-ex]} \BibitemShut
  {NoStop}%
%%CITATION = ARXIV:1806.05010;%%
\bibitem [{\citenamefont {Rout}\ \emph {et~al.}(2017)\citenamefont {Rout},
  \citenamefont {Masud},\ and\ \citenamefont {Mehta}}]{Rout:2017udo}%
  \BibitemOpen
  \bibfield  {author} {\bibinfo {author} {\bibfnamefont {J.}~\bibnamefont
  {Rout}}, \bibinfo {author} {\bibfnamefont {M.}~\bibnamefont {Masud}}, \ and\
  \bibinfo {author} {\bibfnamefont {P.}~\bibnamefont {Mehta}},\ }\href
  {\doibase 10.1103/PhysRevD.95.075035} {\bibfield  {journal} {\bibinfo
  {journal} {Phys. Rev.}\ }\textbf {\bibinfo {volume} {D95}},\ \bibinfo {pages}
  {075035} (\bibinfo {year} {2017})},\ \Eprint
  {http://arxiv.org/abs/1702.02163} {arXiv:1702.02163 [hep-ph]} \BibitemShut
  {NoStop}%
%%CITATION = ARXIV:1702.02163;%%
\bibitem [{\citenamefont {Alonso}\ \emph {et~al.}(2010)\citenamefont {Alonso}
  \emph {et~al.}}]{Alonso:2010fs}%
  \BibitemOpen
  \bibfield  {author} {\bibinfo {author} {\bibfnamefont {J.}~\bibnamefont
  {Alonso}} \emph {et~al.},\ }\href@noop {} {\  (\bibinfo {year} {2010})},\
  \Eprint {http://arxiv.org/abs/1006.0260} {arXiv:1006.0260 [physics.ins-det]}
  \BibitemShut {NoStop}%
%%CITATION = ARXIV:1006.0260;%%
\bibitem [{\citenamefont {Evslin}\ \emph {et~al.}(2016)\citenamefont {Evslin},
  \citenamefont {Ge},\ and\ \citenamefont {Hagiwara}}]{Evslin:2015pya}%
  \BibitemOpen
  \bibfield  {author} {\bibinfo {author} {\bibfnamefont {J.}~\bibnamefont
  {Evslin}}, \bibinfo {author} {\bibfnamefont {S.-F.}\ \bibnamefont {Ge}}, \
  and\ \bibinfo {author} {\bibfnamefont {K.}~\bibnamefont {Hagiwara}},\ }\href
  {\doibase 10.1007/JHEP02(2016)137} {\bibfield  {journal} {\bibinfo  {journal}
  {JHEP}\ }\textbf {\bibinfo {volume} {02}},\ \bibinfo {pages} {137} (\bibinfo
  {year} {2016})},\ \Eprint {http://arxiv.org/abs/1506.05023} {arXiv:1506.05023
  [hep-ph]} \BibitemShut {NoStop}%
%%CITATION = ARXIV:1506.05023;%%
\bibitem [{\citenamefont {Agarwalla}\ \emph {et~al.}(2011)\citenamefont
  {Agarwalla}, \citenamefont {Huber}, \citenamefont {Link},\ and\ \citenamefont
  {Mohapatra}}]{Agarwalla:2010nn}%
  \BibitemOpen
  \bibfield  {author} {\bibinfo {author} {\bibfnamefont {S.~K.}\ \bibnamefont
  {Agarwalla}}, \bibinfo {author} {\bibfnamefont {P.}~\bibnamefont {Huber}},
  \bibinfo {author} {\bibfnamefont {J.~M.}\ \bibnamefont {Link}}, \ and\
  \bibinfo {author} {\bibfnamefont {D.}~\bibnamefont {Mohapatra}},\ }\href
  {\doibase 10.1007/JHEP04(2011)099} {\bibfield  {journal} {\bibinfo  {journal}
  {JHEP}\ }\textbf {\bibinfo {volume} {04}},\ \bibinfo {pages} {099} (\bibinfo
  {year} {2011})},\ \Eprint {http://arxiv.org/abs/1005.4055} {arXiv:1005.4055
  [hep-ph]} \BibitemShut {NoStop}%
%%CITATION = ARXIV:1005.4055;%%
\bibitem [{\citenamefont {Cao}\ \emph {et~al.}(2014)\citenamefont {Cao} \emph
  {et~al.}}]{Cao:2014bea}%
  \BibitemOpen
  \bibfield  {author} {\bibinfo {author} {\bibfnamefont {J.}~\bibnamefont
  {Cao}} \emph {et~al.},\ }\href {\doibase 10.1103/PhysRevSTAB.17.090101}
  {\bibfield  {journal} {\bibinfo  {journal} {Phys. Rev. ST Accel. Beams}\
  }\textbf {\bibinfo {volume} {17}},\ \bibinfo {pages} {090101} (\bibinfo
  {year} {2014})},\ \Eprint {http://arxiv.org/abs/1401.8125} {arXiv:1401.8125
  [physics.acc-ph]} \BibitemShut {NoStop}%
%%CITATION = ARXIV:1401.8125;%%
\bibitem [{\citenamefont {Agarwalla}\ \emph {et~al.}(2017)\citenamefont
  {Agarwalla}, \citenamefont {Ghosh},\ and\ \citenamefont
  {Raut}}]{Agarwalla:2017nld}%
  \BibitemOpen
  \bibfield  {author} {\bibinfo {author} {\bibfnamefont {S.~K.}\ \bibnamefont
  {Agarwalla}}, \bibinfo {author} {\bibfnamefont {M.}~\bibnamefont {Ghosh}}, \
  and\ \bibinfo {author} {\bibfnamefont {S.~K.}\ \bibnamefont {Raut}},\ }\href
  {\doibase 10.1007/JHEP05(2017)115} {\bibfield  {journal} {\bibinfo  {journal}
  {JHEP}\ }\textbf {\bibinfo {volume} {05}},\ \bibinfo {pages} {115} (\bibinfo
  {year} {2017})},\ \Eprint {http://arxiv.org/abs/1704.06116} {arXiv:1704.06116
  [hep-ph]} \BibitemShut {NoStop}%
%%CITATION = ARXIV:1704.06116;%%
\bibitem [{\citenamefont {Ge}\ \emph {et~al.}(2017)\citenamefont {Ge},
  \citenamefont {Pasquini}, \citenamefont {Tortola},\ and\ \citenamefont
  {Valle}}]{Ge:2016xya}%
  \BibitemOpen
  \bibfield  {author} {\bibinfo {author} {\bibfnamefont {S.-F.}\ \bibnamefont
  {Ge}}, \bibinfo {author} {\bibfnamefont {P.}~\bibnamefont {Pasquini}},
  \bibinfo {author} {\bibfnamefont {M.}~\bibnamefont {Tortola}}, \ and\
  \bibinfo {author} {\bibfnamefont {J.~W.~F.}\ \bibnamefont {Valle}},\ }\href
  {\doibase 10.1103/PhysRevD.95.033005} {\bibfield  {journal} {\bibinfo
  {journal} {Phys. Rev.}\ }\textbf {\bibinfo {volume} {D95}},\ \bibinfo {pages}
  {033005} (\bibinfo {year} {2017})},\ \Eprint
  {http://arxiv.org/abs/1605.01670} {arXiv:1605.01670 [hep-ph]} \BibitemShut
  {NoStop}%
%%CITATION = ARXIV:1605.01670;%%
\bibitem [{\citenamefont {Ge}\ and\ \citenamefont
  {Smirnov}(2016)}]{Ge:2016dlx}%
  \BibitemOpen
  \bibfield  {author} {\bibinfo {author} {\bibfnamefont {S.-F.}\ \bibnamefont
  {Ge}}\ and\ \bibinfo {author} {\bibfnamefont {A.~{\relax Yu}.}\ \bibnamefont
  {Smirnov}},\ }\href {\doibase 10.1007/JHEP10(2016)138} {\bibfield  {journal}
  {\bibinfo  {journal} {JHEP}\ }\textbf {\bibinfo {volume} {10}},\ \bibinfo
  {pages} {138} (\bibinfo {year} {2016})},\ \Eprint
  {http://arxiv.org/abs/1607.08513} {arXiv:1607.08513 [hep-ph]} \BibitemShut
  {NoStop}%
%%CITATION = ARXIV:1607.08513;%%
\bibitem [{\citenamefont {Ge}\ and\ \citenamefont {Parke}(2019)}]{Ge:2018uhz}%
  \BibitemOpen
  \bibfield  {author} {\bibinfo {author} {\bibfnamefont {S.-F.}\ \bibnamefont
  {Ge}}\ and\ \bibinfo {author} {\bibfnamefont {S.~J.}\ \bibnamefont {Parke}},\
  }\href {\doibase 10.1103/PhysRevLett.122.211801} {\bibfield  {journal}
  {\bibinfo  {journal} {Phys. Rev. Lett.}\ }\textbf {\bibinfo {volume} {122}},\
  \bibinfo {pages} {211801} (\bibinfo {year} {2019})},\ \Eprint
  {http://arxiv.org/abs/1812.08376} {arXiv:1812.08376 [hep-ph]} \BibitemShut
  {NoStop}%
%%CITATION = ARXIV:1812.08376;%%
\bibitem [{\citenamefont {Tang}\ and\ \citenamefont
  {Zhang}(2018)}]{Tang:2017qen}%
  \BibitemOpen
  \bibfield  {author} {\bibinfo {author} {\bibfnamefont {J.}~\bibnamefont
  {Tang}}\ and\ \bibinfo {author} {\bibfnamefont {Y.}~\bibnamefont {Zhang}},\
  }\href {\doibase 10.1103/PhysRevD.97.035018} {\bibfield  {journal} {\bibinfo
  {journal} {Phys. Rev.}\ }\textbf {\bibinfo {volume} {D97}},\ \bibinfo {pages}
  {035018} (\bibinfo {year} {2018})},\ \Eprint
  {http://arxiv.org/abs/1705.09500} {arXiv:1705.09500 [hep-ph]} \BibitemShut
  {NoStop}%
%%CITATION = ARXIV:1705.09500;%%
\bibitem [{\citenamefont {Abe}\ \emph {et~al.}(2018)\citenamefont {Abe} \emph
  {et~al.}}]{Abe:2016ero}%
  \BibitemOpen
  \bibfield  {author} {\bibinfo {author} {\bibfnamefont {K.}~\bibnamefont
  {Abe}} \emph {et~al.} (\bibinfo {collaboration} {Hyper-Kamiokande}),\ }\href
  {\doibase 10.1093/ptep/pty044} {\bibfield  {journal} {\bibinfo  {journal}
  {PTEP}\ }\textbf {\bibinfo {volume} {2018}},\ \bibinfo {pages} {063C01}
  (\bibinfo {year} {2018})},\ \Eprint {http://arxiv.org/abs/1611.06118}
  {arXiv:1611.06118 [hep-ex]} \BibitemShut {NoStop}%
%%CITATION = ARXIV:1611.06118;%%
\bibitem [{\citenamefont {Ghosh}\ \emph {et~al.}(2016)\citenamefont {Ghosh},
  \citenamefont {Ghoshal}, \citenamefont {Goswami}, \citenamefont {Nath},\ and\
  \citenamefont {Raut}}]{Ghosh:2015ena}%
  \BibitemOpen
  \bibfield  {author} {\bibinfo {author} {\bibfnamefont {M.}~\bibnamefont
  {Ghosh}}, \bibinfo {author} {\bibfnamefont {P.}~\bibnamefont {Ghoshal}},
  \bibinfo {author} {\bibfnamefont {S.}~\bibnamefont {Goswami}}, \bibinfo
  {author} {\bibfnamefont {N.}~\bibnamefont {Nath}}, \ and\ \bibinfo {author}
  {\bibfnamefont {S.~K.}\ \bibnamefont {Raut}},\ }\href {\doibase
  10.1103/PhysRevD.93.013013} {\bibfield  {journal} {\bibinfo  {journal} {Phys.
  Rev.}\ }\textbf {\bibinfo {volume} {D93}},\ \bibinfo {pages} {013013}
  (\bibinfo {year} {2016})},\ \Eprint {http://arxiv.org/abs/1504.06283}
  {arXiv:1504.06283 [hep-ph]} \BibitemShut {NoStop}%
%%CITATION = ARXIV:1504.06283;%%
\bibitem [{\citenamefont {Antusch}\ \emph {et~al.}(2006)\citenamefont
  {Antusch}, \citenamefont {Biggio}, \citenamefont {Fernandez-Martinez},
  \citenamefont {Gavela},\ and\ \citenamefont {Lopez-Pavon}}]{Antusch:2006vwa}%
  \BibitemOpen
  \bibfield  {author} {\bibinfo {author} {\bibfnamefont {S.}~\bibnamefont
  {Antusch}}, \bibinfo {author} {\bibfnamefont {C.}~\bibnamefont {Biggio}},
  \bibinfo {author} {\bibfnamefont {E.}~\bibnamefont {Fernandez-Martinez}},
  \bibinfo {author} {\bibfnamefont {M.~B.}\ \bibnamefont {Gavela}}, \ and\
  \bibinfo {author} {\bibfnamefont {J.}~\bibnamefont {Lopez-Pavon}},\ }\href
  {\doibase 10.1088/1126-6708/2006/10/084} {\bibfield  {journal} {\bibinfo
  {journal} {JHEP}\ }\textbf {\bibinfo {volume} {10}},\ \bibinfo {pages} {084}
  (\bibinfo {year} {2006})},\ \Eprint {http://arxiv.org/abs/hep-ph/0607020}
  {arXiv:hep-ph/0607020 [hep-ph]} \BibitemShut {NoStop}%
%%CITATION = HEP-PH/0607020;%%
\bibitem [{\citenamefont {Eitel}(2001)}]{Eitel:2000by}%
  \BibitemOpen
  \bibfield  {author} {\bibinfo {author} {\bibfnamefont {K.}~\bibnamefont
  {Eitel}} (\bibinfo {collaboration} {KARMEN}),\ }\bibfield  {booktitle} {\emph
  {\bibinfo {booktitle} {{Neutrino physics and astrophysics. Proceedings, 19th
  International Conference, Neutrino 2000, Sudbury, Canada, June 16-21,
  2000}}},\ }\href {\doibase 10.1016/S0920-5632(00)00940-3} {\bibfield
  {journal} {\bibinfo  {journal} {Nucl. Phys. Proc. Suppl.}\ }\textbf {\bibinfo
  {volume} {91}},\ \bibinfo {pages} {191} (\bibinfo {year} {2001})},\ \bibinfo
  {note} {[,191(2000)]},\ \Eprint {http://arxiv.org/abs/hep-ex/0008002}
  {arXiv:hep-ex/0008002 [hep-ex]} \BibitemShut {NoStop}%
%%CITATION = HEP-EX/0008002;%%
\bibitem [{\citenamefont {Astier}\ \emph {et~al.}(2003)\citenamefont {Astier}
  \emph {et~al.}}]{Astier:2003gs}%
  \BibitemOpen
  \bibfield  {author} {\bibinfo {author} {\bibfnamefont {P.}~\bibnamefont
  {Astier}} \emph {et~al.} (\bibinfo {collaboration} {NOMAD}),\ }\href
  {\doibase 10.1016/j.physletb.2003.07.029} {\bibfield  {journal} {\bibinfo
  {journal} {Phys. Lett.}\ }\textbf {\bibinfo {volume} {B570}},\ \bibinfo
  {pages} {19} (\bibinfo {year} {2003})},\ \Eprint
  {http://arxiv.org/abs/hep-ex/0306037} {arXiv:hep-ex/0306037 [hep-ex]}
  \BibitemShut {NoStop}%
%%CITATION = HEP-EX/0306037;%%
\bibitem [{\citenamefont {Huber}\ \emph {et~al.}(2005)\citenamefont {Huber},
  \citenamefont {Lindner},\ and\ \citenamefont {Winter}}]{Huber:2004ka}%
  \BibitemOpen
  \bibfield  {author} {\bibinfo {author} {\bibfnamefont {P.}~\bibnamefont
  {Huber}}, \bibinfo {author} {\bibfnamefont {M.}~\bibnamefont {Lindner}}, \
  and\ \bibinfo {author} {\bibfnamefont {W.}~\bibnamefont {Winter}},\ }\href
  {\doibase 10.1016/j.cpc.2005.01.003} {\bibfield  {journal} {\bibinfo
  {journal} {Comput. Phys. Commun.}\ }\textbf {\bibinfo {volume} {167}},\
  \bibinfo {pages} {195} (\bibinfo {year} {2005})},\ \Eprint
  {http://arxiv.org/abs/hep-ph/0407333} {arXiv:hep-ph/0407333 [hep-ph]}
  \BibitemShut {NoStop}%
%%CITATION = HEP-PH/0407333;%%
\bibitem [{\citenamefont {Kopp}(2008)}]{Kopp:2006wp}%
  \BibitemOpen
  \bibfield  {author} {\bibinfo {author} {\bibfnamefont {J.}~\bibnamefont
  {Kopp}},\ }\href {\doibase 10.1142/S0129183108012303} {\bibfield  {journal}
  {\bibinfo  {journal} {Int. J. Mod. Phys.}\ }\textbf {\bibinfo {volume}
  {C19}},\ \bibinfo {pages} {523} (\bibinfo {year} {2008})},\ \Eprint
  {http://arxiv.org/abs/physics/0610206} {arXiv:physics/0610206 [physics]}
  \BibitemShut {NoStop}%
%%CITATION = PHYSICS/0610206;%%
\bibitem [{\citenamefont {Ghosh}\ \emph {et~al.}(2017)\citenamefont {Ghosh},
  \citenamefont {Goswami},\ and\ \citenamefont {Raut}}]{Ghosh:2014zea}%
  \BibitemOpen
  \bibfield  {author} {\bibinfo {author} {\bibfnamefont {M.}~\bibnamefont
  {Ghosh}}, \bibinfo {author} {\bibfnamefont {S.}~\bibnamefont {Goswami}}, \
  and\ \bibinfo {author} {\bibfnamefont {S.~K.}\ \bibnamefont {Raut}},\ }\href
  {\doibase 10.1142/S0217732317500341} {\bibfield  {journal} {\bibinfo
  {journal} {Mod. Phys. Lett.}\ }\textbf {\bibinfo {volume} {A32}},\ \bibinfo
  {pages} {1750034} (\bibinfo {year} {2017})},\ \Eprint
  {http://arxiv.org/abs/1409.5046} {arXiv:1409.5046 [hep-ph]} \BibitemShut
  {NoStop}%
%%CITATION = ARXIV:1409.5046;%%
\end{thebibliography}%
\end{document}